\documentclass[a4paper,11pt]{article}
\pdfoutput=1 

\usepackage{jheppub} 

\usepackage[T1]{fontenc} 
\usepackage{amsmath}
\usepackage{graphicx,epsfig,wrapfig}

\usepackage{amsthm,mathrsfs}
\usepackage{makeidx}
\usepackage{amsfonts}
\usepackage{amssymb}
\usepackage{amsmath}
\usepackage{mathtools}
\usepackage{fixmath}
\usepackage{amsbsy}
\usepackage[capitalize]{cleveref}
\usepackage{mathrsfs}
\usepackage{slashed}
\usepackage{bm}

\usepackage[normalem]{ulem}
\usepackage{simplewick}

\usepackage{multirow}
\usepackage{braket}

\newcommand{\ta}{\left(}

\newcommand{\tc}{\right)}

\newcommand{\vet}[1]{\bm{{#1}}}

\newcommand{\duv}{\Delta_{\text{\tiny{UV}}}}
\newcommand{\UV}{\text{\tiny{UV}}}
\newcommand{\IR}{\text{\tiny{IR}}}

\newcommand{\MS}{{\overline{\text{MS}}}}
\newcommand{\DS}{{\text{D}}}

\newcommand{\sref}[1]{Sec.~\ref{#1}}

\title{\boldmath Mass sum rules of the electron in quantum electrodynamics}

\author[a,b]{S. Rodini}
\author[c]{A. Metz}
\author[a,b,1]{B. Pasquini,\note{Corresponding author.}}

\affiliation[a]{Dipartimento di Fisica, Universit\`a degli Studi di Pavia,
               27100 Pavia, Italy}
\affiliation[b]{Istituto Nazionale di Fisica Nucleare, Sezione di Pavia,
               27100 Pavia, Italy}
\affiliation[c]{Department of Physics, SERC,
             Temple University, Philadelphia, PA 19122, USA}

\emailAdd{simone.rodini01@universitadipavia.it}
\emailAdd{metza@temple.edu}
\emailAdd{barbara.pasquini@unipv.it}

\abstract{Different decompositions of the nucleon mass, in terms of the masses and energies of the underlying constituents, have been proposed in the literature.
We explore the corresponding sum rules in quantum electrodynamics for an electron at one-loop order in perturbation theory.
To this aim we compute the form factors of the energy-momentum tensor, by paying particular attention to the renormalization of ultraviolet divergences, operator mixing and scheme dependence.
We clarify the  expressions of all the proposed sum rules in the electron rest frame in terms of renormalized operators. 
Furthermore, we consider  the same sum rules in a moving frame, where they become energy decompositions. 
Finally, we discuss some implications of our study on the mass sum rules for the nucleon.}

\begin{document} 
\maketitle
\flushbottom

\section{Introduction}
Understanding the internal structure of hadrons --- most notably the nucleon --- is one of the longstanding problems in quantum chromodynamics (QCD), where the ultimate goal is a complete description of the hadron structure in terms of quarks and gluons, the fundamental degrees of freedom of QCD. 
For example, various types of (multi-dimensional) parton distribution functions (PDFs) encode important aspects of the hadron structure
(see, e.g.,~\cite{Diehl:2015uka,Bacchetta:2016ccz,Belitsky:2003nz,Meissner:2009ww,Lorce:2011dv} and references therein).
At energy scales that are comparable with hadron masses, the PDFs are entirely non-perturbative objects, which means that perturbative QCD cannot provide any reliable insights into PDFs in this kinematical regime.
Therefore, constraints on PDFs are coming exclusively from high-energy scattering experiments, numerical calculations in lattice QCD, or (non-perturbative) models of hadrons. 
 
A special role is played by {\it global} properties of hadrons such as their charges, spin and mass. 
Some of these global properties can be obtained by performing suitable integrals of PDFs, and hence we need to know them over the complete range of integration. 
This point is related to the fact that PDFs are defined through matrix elements of non-local operators, while global properties of hadrons are related to local operators.
In this context, the energy-momentum tensor (EMT) has attracted a lot of attention recently.
For (almost) all the available definitions, the EMT is given by a local operator whose matrix elements are parametrized in terms of form factors, which give access to the spin, the mass, and the pressure and shear distributions of hadrons~\cite{Ji:1996ek,Polyakov:2002yz,Polyakov:2018zvc,Lorce:2015lna,Lorce:2018egm}, and as such contain a wealth of information. 
The direct extraction of the EMT form factors from experiment is challenging, although first proof-of-principle studies exist~\cite{Burkert:2018bqq,Kumericki:2019ddg}.
Calculations of these form factors have been performed in different models (see, e.g., Refs.~\cite{Polyakov:2018zvc,Lorce:2018egm} and references therein)
and in lattice QCD~\cite{Hagler:2003jd,Gockeler:2003jfa,Hagler:2007xi,Yang:2018nqn,Yang:2014xsa,Shanahan:2018nnv}. 

A lot of work has already been done for what concerns the spin decomposition of the nucleon, as well as the pressure and shear distributions (see, e.g., ~\cite{Leader:2013jra,Lorce:2017wkb,Burkert:2018bqq,Polyakov:2018zvc} and references therein). 
In the present study, we will focus on the global property of the mass. 
Understanding the internal structure of hadrons is intimately related to understanding the origin of its mass from the mass and energies of the partons. 
Different mass decompositions (sum rules) related to the EMT have been proposed in the literature~\cite{Ji:1994av,Ji:1995sv,Roberts:2016vyn,Lorce:2017xzd,Hatta:2018sqd}. 
It has also been argued that photo- and electro-production of quarkonia close to the kinematical threshold can add to the understanding of the nucleon mass~\cite{Kharzeev:1995ij,Joosten:2018gyo,Hatta:2018ina,Ali:2019lzf,Hatta:2019lxo,Mamo:2019mka,Wang:2019mza}. 

Here we explore the proposed mass decompositions  by Ji~\cite{Ji:1994av}, Lorc\'e~\cite{Lorce:2017xzd}, and Hatta, Rajan, Tanaka~\cite{Hatta:2018sqd} for an electron in quantum electrodynamics (QED) by using perturbation theory up to one-loop order.
Normally, it is not common to talk about a decomposition of the electron mass in QED.
One rather just distinguishes between the physical (measurable) mass and the bare mass when renormalizing the theory, where the renormalized fermion propagator has a pole at the physical electron mass.
However, the physical electron can be seen as a dressed particle surrounded by a cloud of (virtual) photons, electrons, and positrons, which may be interpreted as (constituent) ``partons'' contained in the physical electron, providing a close analogy to the partonic structure of hadrons~\cite{Bacchetta:2015qka,Miller:2014vla,Hoyer:2009sg,Brodsky:2000ii}.
The mass sum rules presented in Refs.~\cite{Ji:1994av,Lorce:2017xzd,Hatta:2018sqd} allow one to identify separate contributions from the constituents to the mass of the physical electron. 
We perform the calculation to first order in the fine structure constant $\alpha$, which corresponds to considering quantum fluctuations of the physical electron into a photon and an electron. 
Up to this order, the topologies of the QED diagrams for the EMT are the same as for a quark target in perturbative QCD, so that the results in the two cases basically just differ by a color factor.
One-loop QED results for the total EMT of the electron are available in the literature~\cite{Berends:1975ah,Milton:1976jr,Milton:1977je}, but these works do not distinguish between the individual contributions from the electron and photon constituents.
To the best of our knowledge, in the forward limit the separate contributions to the EMT form factors have been discussed for the first time in Ref.~\cite{Ji:1998bf}.
Here we revisit this work, which explores the mass sum rule of Ref.~\cite{Ji:1994av}, by paying specific attention to the proper renormalization of the form factors.
The renormalization involves operator mixing and, of course, leads to scheme-dependent results.
We also investigate the two mass decompositions that were suggested more recently in Refs.~\cite{Lorce:2017xzd,Hatta:2018sqd}. 
We discuss a (new) renormalization scheme for the EMT, including its potential relevance for the sum rule of Ref.~\cite{Hatta:2018sqd}.
Our analysis furthermore suggests that the sum rule presented in Ref.~\cite{Ji:1994av} should be modified somewhat.
Moreover, we identify renormalized operators in relation to the sum rule of Ref.~\cite{Lorce:2017xzd}.
In particular, we find that the four-term decompositions in both Ref.~\cite{Ji:1994av} and Ref.~\cite{Lorce:2017xzd} actually have three non-trivial terms only.
We note in passing that the spin decomposition of a dressed electron in QED has been studied intensively in a number of papers 
(see, e.g., Refs.~\cite{Harindranath:1998ve,Brodsky:2000ii,Burkardt:2008ua,Kanazawa:2014nha,Liu:2014fxa,Ji:2015sio}).
Therefore, it seems timely to take a (fresh) look at the corresponding problem for the electron mass.

The paper is organized as follows: In~\sref{sec_definitions}, we review the basic properties of the EMT and give the parametrization of the EMT matrix elements in terms of form factors, while in~\sref{sec_renormalization} we discuss the renormalization procedure leading to the renormalized EMT form factors. 
In~\sref{sec_decompositions}, we examine the three aforementioned mass sum rules available in the literature, and we consider those sum rules defined in  the electron rest frame also in a moving frame, where they become energy decompositions.
We summarize our results in~\sref{sec_conclusions}.

\section{Definitions}
\label{sec_definitions}
The ``canonical'' EMT is defined as the Noether current associated with the space-time translational invariance of the Lagrangian, and therefore it satisfies the continuity equation
\begin{equation}
\partial_\mu T_C^{\mu\nu}  = 0.\label{continuity_Eq}
\end{equation}
As  known, we have the freedom to modify the expression of the EMT by adding a superpotential,
\begin{equation}
T^{\mu\nu} = T_C^{\mu\nu} + \partial_\rho \Phi^{\rho\mu\nu},
\end{equation}
with $\Phi^{\rho\mu\nu} = - \Phi^{\mu\rho\nu}$. 
This corresponds to the Belinfante-Rosenfeld~\cite{Belinfante1,Belinfante2,Rosenfeld} procedure that allows one to incorporate specific properties into the EMT, such as the symmetry in the Lorentz indices and the gauge invariance. 
Having a symmetric EMT is not essential though in quantum field theory, as the antisymmetric part of the EMT is associated with the spin of the particles. 
But the gauge invariance is an essential property for the EMT. 
Contrary to popular belief, it is possible to derive a symmetric and gauge invariant EMT via the ``canonical'' technique of the Noether current without resorting to Belinfante's symmetrization technique, as shown in Refs.~\cite{Montesinos:2006th,Eriksen:1979vq,Takahashi:1985dt,Munoz:1996wp}.
Since the antisymmetric part of the EMT does not contribute to the forward limit in which we are interested, we will use the symmetric form, i.e.
\begin{align}
T_e^{\mu\nu} &= Z_2\ \bar \psi \frac{i}{4}\gamma^{\{\mu}\overset{\leftrightarrow}{\partial^{\nu\}}}  \psi - Z_2 \mu^{2\varepsilon}\ e\bar \psi \gamma^{\{\mu} A^{\nu\}} \psi, \label{Te}\\
T_\gamma^{\mu\nu} &= -Z_3\ F^{\mu\alpha}F^\nu_{\ \alpha} + Z_3\ \frac{g^{\mu\nu}}{4}F^{\alpha\beta}F_{\alpha\beta}, \label{Tgamma}\\
T^{\mu\nu} &= T_e^{\mu\nu} + T_\gamma^{\mu\nu},
\end{align}
where $a^{\{\mu}b^{\nu\}}=a^\mu b^\nu+a^\nu b^\mu$  for any tensor. 
The indices $e$ and $\gamma$ refer to the separate electron and photon contributions, respectively.
In Eqs.~\eqref{Te} and~\eqref{Tgamma}, all the fields and the elementary charge $e$ are renormalized, with $Z_i = 1+\delta_i$ denoting the standard Lagrangian counterterms.
We have used dimensional regularization in $d=4-2\varepsilon$ dimensions with the mass scale $\mu$.

The forward matrix element of the EMT is parametrized in terms of  form factors as~\cite{Ji:1996ek}
\begin{align}
\braket{e(P)|T^{\mu\nu}_i|e(P)} &\equiv \braket{T^{\mu\nu}_i}  =  2P^{\mu}P^{\nu}A_i(0) + 2m^2g^{\mu\nu}\bar C_i(0)  \notag \\
& = \ta 2P^\mu P^\nu - \frac{g^{\mu\nu}}{2}m^2\tc A_i(0) + \frac{g^{\mu\nu}}{2} m^2\ta A_i(0) + 4\bar C_i(0)\tc  ,
\label{GeneralEMTParametrization}
\end{align}
where $m$ is  the electron mass. In Eq.~\eqref{GeneralEMTParametrization}, $A_i(0)$ and $\bar C_i(0)$ ($i=e,\gamma)$ are the EMT form factors, calculated at zero-momentum transfer ($\Delta = 0$): the $A_i(0)$ form factors are associated with the traceless part of the EMT, while the trace of the EMT is given by the combination $A_i(0) + 4\bar C_i(0)$.
The electron and photon form factors are not independent, since the conservation of the total EMT imposes the sum rules
\begin{equation}
A_e(0)+A_\gamma(0) = 1, \quad \bar C_e(0)+\bar C_\gamma(0)  = 0.
\label{veryFundamentalSumRules}
\end{equation}
\begin{figure}[t]
\centering
{\includegraphics[width=.2\textwidth]{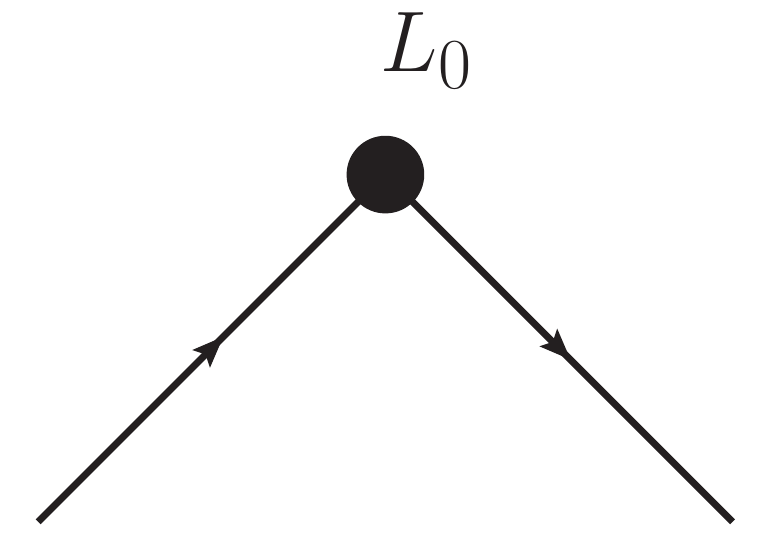}} \quad
{\includegraphics[width=.2\textwidth]{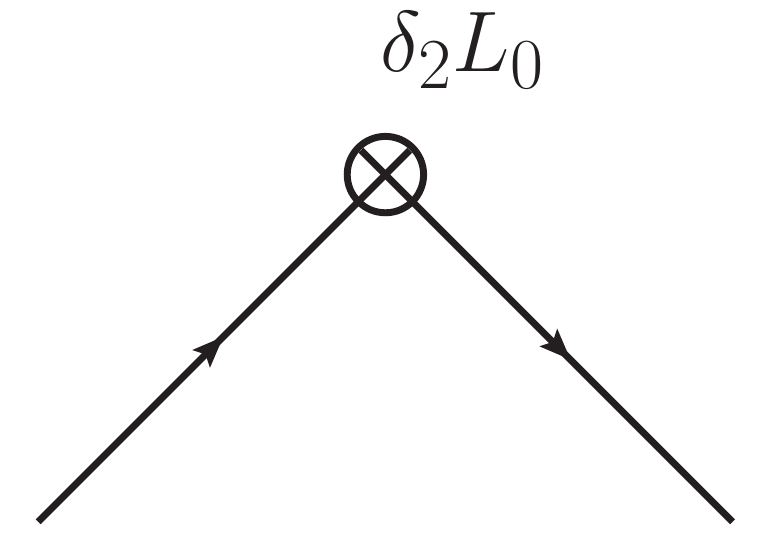}}\\
\vspace{0.7 truecm}
{\includegraphics[width=.2\textwidth]{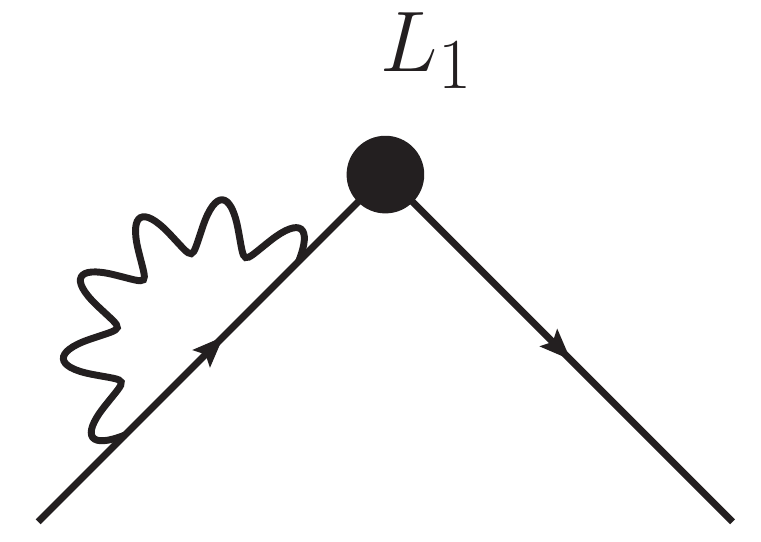}} \quad
{\includegraphics[width=.2\textwidth]{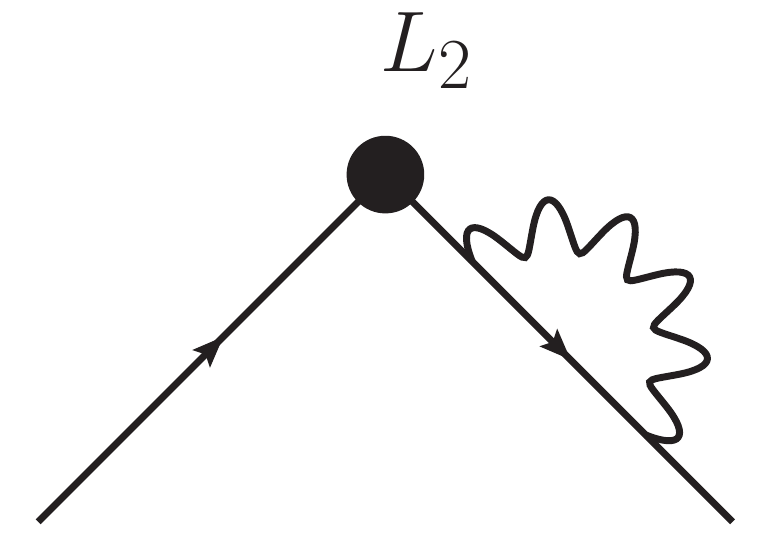}} \quad
{\includegraphics[width=.2\textwidth]{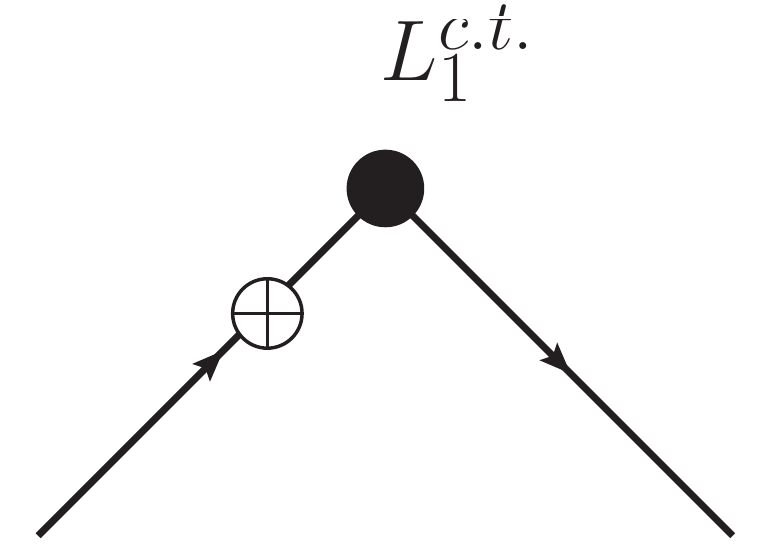}} \quad
{\includegraphics[width=.2\textwidth]{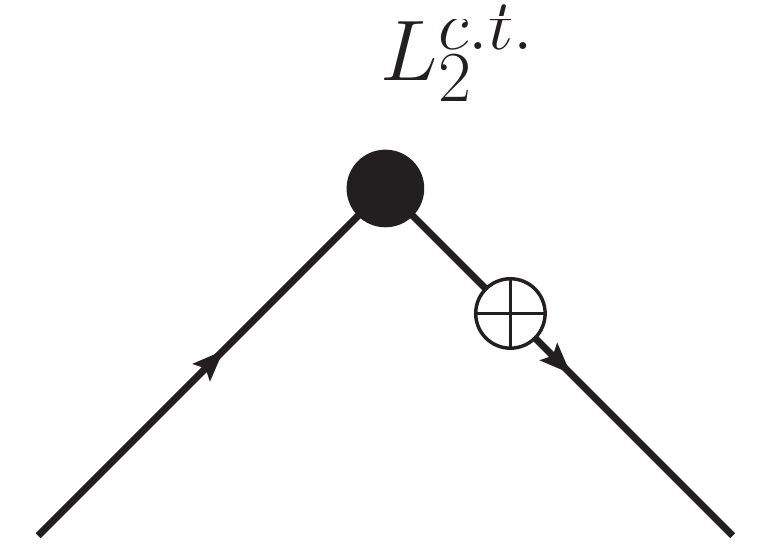}}  \quad\\
\vspace{0.7 truecm}
{\includegraphics[width=.2\textwidth]{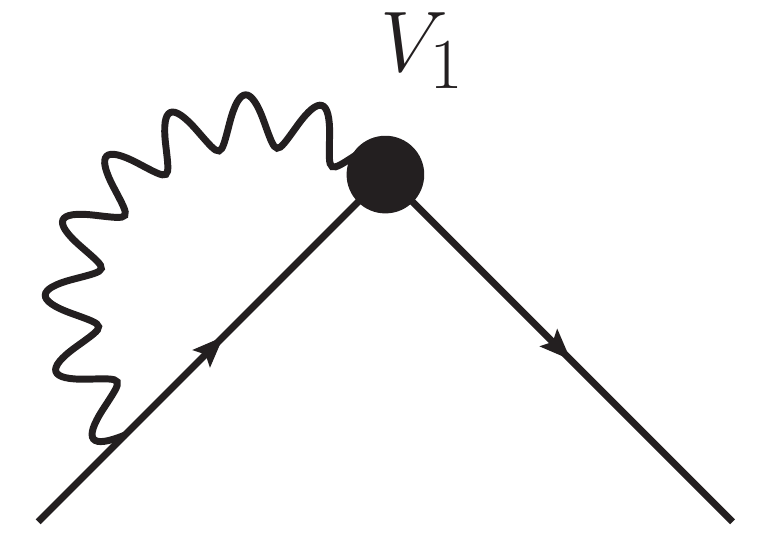}} \quad
{\includegraphics[width=.2\textwidth]{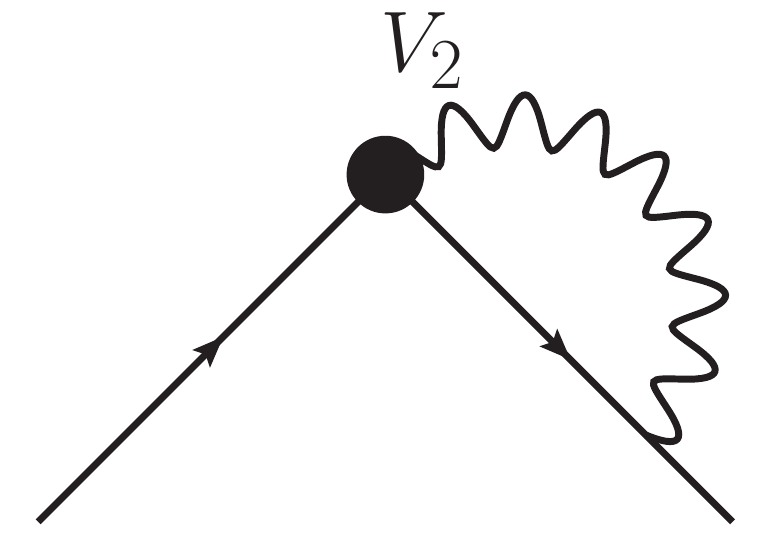}} \quad
{\includegraphics[width=.2\textwidth]{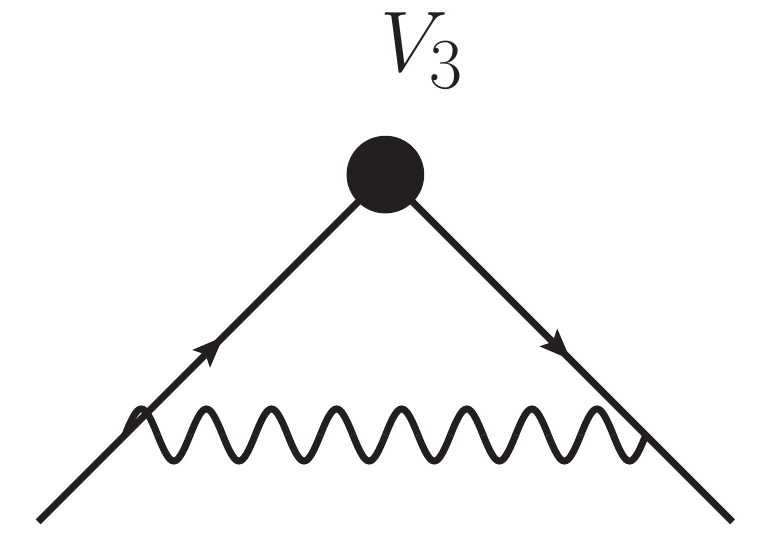}}
{\includegraphics[width=.2\textwidth]{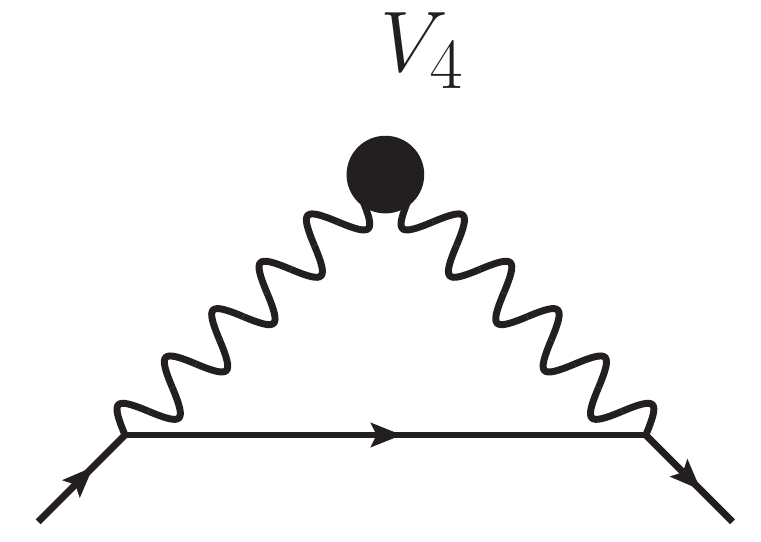}} 
\caption{Relevant diagrams for the calculation of the electron EMT at $O(\alpha)$. 
We use the common notation of a crossed dot for the counterterm diagrams and a black solid dot to indicate the EMT insertion into the Green function.
See text for more details.}
\label{FeyDiag}
\end{figure}

The EMT matrix element can be calculated at any order in $\alpha$ from  the Green function with the insertion of the EMT operator, i.e.
\begin{equation}
\langle{\rm T}\left[ T_i^{\mu\nu}(0)\exp\left( i\int dx\mathcal{L}_I\right)\right]\rangle, \quad \text{ with } \mathcal{L}_I = -e\bar \psi \slashed{A}\psi,
\label{emt_matrix_element}
\end{equation}
where ${\rm T}$ indicates the time-ordered product. 
The space-time point at which the EMT is evaluated is irrelevant thanks to the translational invariance of the forward matrix element. 
In Fig.~\ref{FeyDiag}, we illustrate the diagrams associated with the expansion of Eq.~\eqref{emt_matrix_element} up to $O(\alpha)$. 
$L_0$ is the diagram corresponding to tree level contribution, and $\delta_2 L_0$ is the overall vertex counterterm.  
Since the total EMT is renormalized with the standard Lagrangian renormalization and we are just considering the matrix elements for an electron state, the vertex counterterm coincides with the counterterm for the electron field.
$L_{1,2}$ are the diagrams with the leg-loop corrections, while $L_{1,2}^{c.t.}$ give the corresponding counterterms.
$V_{1,2}$ are the diagrams associated with the interaction term present in $T_e^{\mu\nu}$, whereas $V_3$ is the one-loop electron vertex correction that arises from the derivative term in $T_e^{\mu\nu}$.
Finally, $V_4$ is the one-loop vertex correction with the photon coupled directly to the external operator.

\section{Renormalization}
\label{sec_renormalization}
In the calculation of the diagrams in Fig.~\ref{FeyDiag}, one finds ultraviolet (UV) divergences in the separate electron and photon contributions. Moreover, if one splits the electron contribution according to
\[
L_{tot} = (1+\delta_2)L_0 + L_1+L_1^{c.t.}+ L_2+L_2^{c.t.}, \quad V_1+V_2, \quad V_3,
\]
one infrared divergence shows up in $L_{tot}$ and $V_3$ due to the loop integrals, whereas $V_1+V_2$ is infrared safe because of the four-particle vertex where the photon couples directly with the external operator.
However, we are not concerned about the infrared divergences in the individual diagrams, 
as  we are interested in the total electron contribution
$
L_{tot} + V_1 + V_2 + V_3
$
which is infrared safe.
The separate diagrams $L_i$ depend on the renormalization scheme of the Lagrangian, but $L_{tot} $ does not. 
We employ dimensional regularization for both the UV and infrared divergences, with $\varepsilon_{\UV}>0$ and $\varepsilon_{\IR}>0$ the corresponding dimension parameters.
We obtain the following results:
\begin{align}
L_{tot}\ta \Delta = 0\tc & = 2P^\mu P^\nu\ta 1 + \frac{\alpha}{\pi\varepsilon_{\IR}}-\frac{\alpha}{\pi}  - \mathcal{P}\tc   ,
\label{bare1} \\[0.2cm]
(V_1+V_2)\ta \Delta = 0\tc &= 2P^\mu P^\nu\ta -2\mathcal{P}-\frac{3\alpha}{2\pi}\tc  -2m^2g^{\mu\nu}\ta \mathcal{P}+\frac{\alpha}{4\pi}\tc  ,
\\[0.2cm]
V_3\ta \Delta = 0\tc 
& =2P^\mu P^\nu  \ta -\frac{\alpha}{\pi\varepsilon_{\IR}}+\frac{14\alpha}{9\pi} +\frac{1}{3} \mathcal{P}\tc+ 2m^2g^{\mu\nu} \ta \frac{5}{3}\mathcal{P}+\frac{7\alpha}{36\pi}\tc,
\\[0.2cm]
V_4\ta \Delta = 0\tc & = 2P^\mu P^\nu  \ta \frac{8}{3}\mathcal{P}+\frac{17\alpha}{18\pi}\tc+ 2m^2g^{\mu\nu}  \ta -\frac{2}{3}\mathcal{P} + \frac{\alpha}{18\pi}\tc,\label{bare2}
\end{align} 
where we defined 
\[
\mathcal{P} = \frac{\alpha \Gamma(\varepsilon_{\UV})}{4\pi} \ta \frac{4\pi\mu^2}{m^2} \tc^{\varepsilon_{\UV}} = \frac{\alpha}{4\pi}\ta\duv + \log\ta\frac{\mu^2}{m^2}\tc\tc= \frac{\alpha}{4\pi}\ta\frac{1}{\varepsilon_{\UV}} - \gamma_E + \log\ta4\pi\tc + \mathscr{L}\tc,
\]
with  $
\mathscr{L} = \log\ta \frac{\mu^2}{m^2}
\tc$.
The bare (i.e., with Lagrangian renormalization only) form factors $A_i(0)$ and $\bar C_i(0)$ can be extracted from Eqs.~(\ref{bare1})--(\ref{bare2}) as the coefficients of $2P^{\mu}P^\nu$ and $g^{\mu\nu}$, respectively:
\begin{align}
A_e(0) &= 1 -\frac{8}{3}\mathcal{P}-\frac{17}{18}\frac{\alpha}{\pi},\label{eq-ae-nr}\\
\bar C_e(0) &= \frac{2}{3}\mathcal{P}-\frac{1}{18}\frac{\alpha}{\pi},\label{eq-ce-nr}\\
A_\gamma(0) &= \frac{8}{3}\mathcal{P}+\frac{17}{18}\frac{\alpha}{\pi},\label{eq-ag-nr}\\
\bar C_\gamma(0) &= -\frac{2}{3}\mathcal{P}+\frac{1}{18}\frac{\alpha}{\pi}.\label{eq-cg-nr}
\end{align}
Because of the continuity equation~\eqref{continuity_Eq}, the renormalization of the Lagrangian also renormalizes the total EMT.
But it does not renormalize the UV divergences of the separate contributions from the electron and photon EMT.
Therefore, additional renormalization is required.

The renormalization of the UV divergences of the individual composite operators defining the EMT is a rather tricky subject, discussed in detail in Refs.~\cite{Hatta:2018sqd,Tanaka:2018nae} for QCD.
Following the procedure outlined in these works, we introduce the operators\footnote{To simplify the notation, we omit the tensor indices in the operators ${\cal O}_i$.}
\begin{align}
\mathcal{O}_3 &= Z_2 \frac{i}{4}\bar \psi \gamma^{\{\mu}\overset{\leftrightarrow}{D^{\nu\}}}\psi, \qquad \mathcal{O}_4 = g^{\mu\nu}Z_2Z_m  m\bar \psi\psi,\\
\mathcal{O}_1 &= -Z_3F^{\mu\alpha}F^\nu_{\ \alpha},  \quad\qquad\mathcal{O}_2 = g^{\mu\nu}Z_3F^{\alpha\beta}F_{\alpha\beta},
\end{align}
which allows us to write
\begin{equation}
T^{\mu\nu}  = \mathcal{O}_1 + \frac{\mathcal{O}_2}{4}+ \mathcal{O}_3 .
\end{equation}
We will use the modified minimal subtraction ($\MS$) scheme instead of the MS scheme employed in Refs.~\cite{Hatta:2018sqd,Tanaka:2018nae}. 
The transition between the two schemes is simply performed with the replacement $1/\varepsilon_{\UV} \rightarrow \duv$.
We will carry out the renormalization for the operators evaluated between electron states.
Therefore, some of the counterterms associated with the photon contributions vanish.  
Additional work would be required to renormalize the EMT evaluated between photon states. 
Considering (just) the forward limit does not simplify the renormalization procedure, as the counterterms are independent of the external kinematics.

We repeat that the Lagrangian renormalization for all the fields is understood --- at $O(\alpha)$ the charge renormalization is not present.
For the renormalization of the composite operators ${\cal O}_i$ we consider the following system of equations:
\begin{align}
\mathcal{O}_1^R &= Z_T\mathcal{O}_1+Z_M\mathcal{O}_2+Z_L\mathcal{O}_3+Z_S\mathcal{O}_4, \label{renormOP1}\\
\mathcal{O}_2^R &= Z_F\mathcal{O}_2+Z_C\mathcal{O}_4, \label{renormOP2} \\
\mathcal{O}_3^R &= Z_\psi\mathcal{O}_3+Z_K\mathcal{O}_4+Z_Q\mathcal{O}_1+Z_B\mathcal{O}_2, \\
\mathcal{O}_4^R &= \mathcal{O}_4, \label{renormOP4}
\end{align}
where $\mathcal{O}_4$ is renormalization-invariant. 
Equations~\eqref{renormOP1}--\eqref{renormOP4} are obtained by considering all the independent operators with the same dimension and the same Lorentz structure (second rank tensor in this case). 
The trace part in Eq.~\eqref{renormOP2} is simplified using the well known results for the trace anomaly, see Refs.~\cite{Adler:1976zt,Adler:2004qt,Nielsen:1977sy,Tarrach:1981bi}. 
The result for the trace anomaly also fixes the counterterms of the trace of the photon EMT (in the $\MS$ scheme),
\begin{equation}
Z_F = 1+\frac{\beta(e)}{e}\duv = 1+\frac{\alpha}{3\pi}\duv, \quad Z_C = 2\gamma_m\duv = \frac{3\alpha}{\pi}\duv, 
\label{trace_counterterms}
\end{equation}
where we used the definitions of the QED $\beta$-function and the anomalous dimension of the electron mass at order $\alpha$,
\begin{equation}
\frac{\beta(e)}{2e} = -\frac{\alpha\beta_0}{8\pi}, \quad \beta_0 = -\frac{4}{3},\quad \gamma_m = \frac{3\alpha}{2\pi}.
\end{equation}
The invariance of the total EMT under renormalization imposes the following constraints on the counterterms:
\begin{align}
& Z_T+Z_Q = 1,\label{ct1} \\
& Z_L+Z_\psi = 1,\\
& Z_M+Z_B + \frac{Z_F}{4} = \frac{1}{4},\\
& Z_S+Z_K + \frac{Z_C}{4} = 0. \label{ct4}
\end{align}
We can also define the traceless operators $\widetilde{\mathcal{O}}_i$ for the electron and the photon 
\begin{align}
\widetilde{\mathcal{O}}_1^R &= \mathcal{O}_1^{R}+\frac{1}{4}\ta 1-\frac{\beta(e)}{2e} + x\tc\mathcal{O}_2^R +\frac{y-\gamma_m}{4}\mathcal{O}_4^R,\label{traceless_photon_renorm} \\
\widetilde{\mathcal{O}}_3^{R} &= \mathcal{O}_3^{R}-\frac{x}{4}\mathcal{O}_2^R-\frac{1+y}{4} \mathcal{O}_4^R, \label{treceless_electron_renorm}
\end{align}
where $x,y$ are finite $\alpha$-dependent parameters starting at $O(\alpha)$. 
We recall from Eq.~\eqref{GeneralEMTParametrization} that the traceless operators are directly related to the $A_i(0)$ form factors.
As already pointed out in Ref.~\cite{Tanaka:2018nae}, the Eqs.~(\ref{ct1})--(\ref{ct4}) do not add new constraints on the values of $x,y$. 
To fix these two parameters one may use the $\MS$ scheme in which one requires a vanishing finite part for all the counterterms.
Before elaborating more on this point, we recall that the trace of the renormalized electron and photon operators are, generally, linear combinations of the renormalized traces of the electron and photon operators (see, e.g., Eq.~(9) in Ref.~\cite{Hatta:2019lxo}), 
\begin{align}
\langle\ta T_{e,R}\tc^\mu_\mu\rangle & = (1+y)\langle\ta m \bar\psi \psi\tc_R\rangle + x\langle\ta F^{\mu\nu}F_{\mu\nu}\tc_R\rangle, \label{trace_mixing-1}\\ 
\langle\ta T_{\gamma,R}\tc^\mu_\mu\rangle & = (\gamma_m-y)\langle\ta m \bar\psi \psi\tc_R\rangle + \ta\frac{\beta(e)}{2e} - x\tc\langle\ta F^{\mu\nu}F_{\mu\nu}\tc_R\rangle. \label{trace_mixing-2}
\end{align}
One may also consider choosing $x, y$ such that this system of equations becomes diagonal.
This is achieved in what we call the ``diagonal"  ($\DS$) scheme. In this scheme, we choose $x = 0$, in order to remove the photon contribution from the r.h.s.~of Eq.~\eqref{trace_mixing-1}, and $y=\gamma_m$, in order to remove the electron contribution  from the r.h.s.~of Eq.~\eqref{trace_mixing-2}.
We emphasize that this definition of the $\DS$ scheme  holds to all orders in perturbation theory and can be used also in QCD.
In the $\MS$ scheme the value of $x$ is determined by the counterterms $Z_T,Z_F$ (see Ref.~\cite{Hatta:2018sqd}). 
Since  the $O(\alpha)$ contribution to $Z_T$ is related to the matrix elements of the EMT between photon states, we cannot derive the full result for $x$. 
However, for our purposes of the one-loop calculation, the  $O(\alpha)$ term of $x$ is not relevant since for an electron state the product $x\langle\ta F^{\mu\nu}F_{\mu\nu}\tc_R\rangle$ is of $O(\alpha^2)$. Therefore, we use the value $x=0$ also in the $\MS$ scheme.
To $O(\alpha)$ we have for the two schemes
\begin{equation}
x = 0, \quad y = \begin{cases}
\frac{\alpha}{3\pi},& \MS \\
\gamma_m=\frac{3\alpha}{2\pi}, & \DS
\end{cases}.
\label{scheme}
\end{equation}

Below we will show one-loop results in both the $\MS$ scheme and the $\DS$ scheme.

The counterterms that involve the parameters $x,y$ are $Z_{B,M,K,S}$ (see Eqs.~(4.21)--(4.28) of Ref.~\cite{Hatta:2018sqd}).
The counterterms $Z_{\psi,Q,L,T}$ are fixed from the evolution equations of the form factors $A_i(0)$. 
Note that the additional renormalization of the form factors $A_i(0)$  is carried out using the same   dimensional regularization introduced for the Lagrangian renormalization. All the scale-dependence of the $A_i(0)$ form factors can come only from the Lagrangian renormalization. Therefore, the scale dependence of the $A_i(0)$ and the $A_i^R(0)$ is the same.
From the results in Eqs.~\eqref{eq-ae-nr}-\eqref{eq-cg-nr}, we can immediately derive
\begin{align}
\frac{\partial}{\partial \ln\mu}A_e(0)= -\frac{\partial}{\partial \ln\mu}A_\gamma(0) = -\frac{4\alpha}{3\pi} A_e(0) .
\end{align} 
The full evolution equations also require knowledge of $A_e$ and $A_\gamma$ for a photon state.
Here we will not consider them since we are dealing with an electron state only. 
However,  the $A_i(0)$ are matrix elements of twist-two, spin-2 electron and photon operators, and follow the evolution equations of the second moment of the Dokshitzer-Gribov-Lipatov-Altarelli-Parisi (DGLAP) evolution equations for the flavor-singlet part of the unpolarized PDFs. 
To one loop-order, one can use the QCD results of Ref.~\cite{Hatta:2018sqd} with  $n_f=1$, $C_F=1$, $C_A=0$ for the QED case.
Following the procedure illustrated in detail in Ref.~\cite{Tanaka:2018nae}, we obtain in the two schemes 
\begin{align}
Z_T &= 1, \qquad\qquad\qquad Z_Q = 0, \qquad\qquad \quad Z_\psi = 1+\frac{2\alpha}{3\pi}\duv, \label{all_counter_i}\\
Z_L &= -\frac{2\alpha}{3\pi}\duv,\qquad\, \,Z_M  =  -\frac{\alpha}{12\pi}\duv,\quad Z_B = 0,\\
Z_S  &=  \begin{cases}
-\frac{7\alpha}{12\pi}\duv, & \MS \\
-\frac{7\alpha}{24\pi} -\frac{7\alpha}{12\pi}\duv,  & \DS
\end{cases},\qquad\quad\quad\,\,\,
Z_K =  \begin{cases}
-\frac{\alpha}{6\pi}\duv,  & \MS \\
\frac{7\alpha}{24\pi}-\frac{\alpha}{6\pi}\duv, & \DS
\end{cases}.
\label{all_counter_f}
\end{align}
The difference between the two schemes is a finite part of $O(\alpha)$.
Using the values for the counterterms in Eqs.~(\ref{trace_counterterms}) and (\ref{all_counter_i})--(\ref{all_counter_f}), along with the (trivial) tree-level results
\begin{equation}
\braket{\mathcal{O}_3}_{tree} = 2P^{\mu}P^\nu, \quad \braket{\mathcal{O}_4}_{tree} = 2m^2g^{\mu\nu}, \quad \braket{\mathcal{O}_{1,2}}_{tree} = 0,
\end{equation}
we obtain from Eqs.~\eqref{renormOP1}--\eqref{renormOP4} in the $\MS$ scheme
\begin{align}
\braket{\mathcal{O}_3^R}^{\MS} &= \braket{\mathcal{O}_3}+\frac{2\alpha}{3\pi}\duv \: (2 P^\mu P^\nu)   -\frac{\alpha}{6\pi}\duv \: (2m^2g^{\mu\nu}) , \\
\braket{\mathcal{O}_1^R}^{\MS} &= \braket{\mathcal{O}_1}-\frac{2\alpha}{3\pi}\duv \: (2 P^\mu P^\nu)    -\frac{7\alpha}{12\pi}\duv \: (2m^2g^{\mu\nu}) , \\
\braket{\mathcal{O}_2^R}^{\MS} &= \braket{\mathcal{O}_2}+ \frac{3\alpha}{\pi}\duv\: (2m^2g^{\mu\nu})  .
\end{align}
The corresponding results with the counterterms in the $\DS$ scheme are
\begin{align}
\braket{\mathcal{O}_3^R}^{\DS} &= \braket{\mathcal{O}_3}+\frac{2\alpha}{3\pi}\duv\: (2 P^\mu P^\nu)   + \ta\frac{7\alpha}{24\pi}-\frac{\alpha}{6\pi}\duv\tc \: (2m^2g^{\mu\nu}) , \\
\braket{\mathcal{O}_1^R}^{\DS} &= \braket{\mathcal{O}_1}-\frac{2\alpha}{3\pi}\duv\: (2 P^\mu P^\nu)   + \ta -\frac{7\alpha}{24\pi}-\frac{7\alpha}{12\pi}\duv\tc \: (2m^2g^{\mu\nu}) , \\
\braket{\mathcal{O}_2^R}^{\DS} &= \braket{\mathcal{O}_2}+ \frac{3\alpha}{\pi}\duv \: (2m^2g^{\mu\nu})  .
\end{align}
As a result, the renormalized expressions for the Feynman diagrams read
\begin{align}
L_{tot}^R\ta \Delta = 0\tc & = 2P^\mu P^\nu \ta 1 + \frac{\alpha}{\pi\varepsilon_{\IR}}-\frac{\alpha}{\pi}  - \frac{\alpha \mathscr{L}}{4\pi}\tc  ,\\[0.2cm]
(V_1+V_2)^R \ta \Delta = 0\tc &= 
\begin{cases}
2P^\mu P^\nu\ta -\frac{\alpha \mathscr{L}}{2\pi} -\frac{3\alpha}{2\pi}\tc   -2m^2g^{\mu\nu} \ta \frac{\alpha \mathscr{L}}{4\pi}+\frac{\alpha}{4\pi}\tc  & \MS, \\
2P^\mu P^\nu\ta -\frac{\alpha \mathscr{L}}{2\pi} -\frac{3\alpha}{2\pi}\tc   -2m^2g^{\mu\nu}\ta \frac{\alpha \mathscr{L}}{4\pi}-\frac{\alpha}{24\pi}\tc  & \DS,
\end{cases} \\[0.2cm]
V_3^R\ta \Delta = 0\tc 
& =2P^\mu P^\nu  \ta -\frac{\alpha}{\pi\varepsilon_{\IR}}+\frac{14\alpha}{9\pi} +\frac{\alpha \mathscr{L}}{12\pi}\tc+ 2m^2g^{\mu\nu} \ta \frac{5\alpha \mathscr{L}}{12\pi}+\frac{7\alpha}{36\pi}\tc,\\[0.2cm]
V_4^R\ta \Delta = 0\tc & = 
\begin{cases}
2P^\mu P^\nu  \ta \frac{2\alpha \mathscr{L}}{3\pi}+\frac{17\alpha}{18\pi}\tc+ 2m^2g^{\mu\nu}  \ta -\frac{\alpha \mathscr{L}}{6\pi}+ \frac{\alpha}{18\pi}\tc & \MS, \\
2P^\mu P^\nu  \ta \frac{2\alpha \mathscr{L}}{3\pi}+\frac{17\alpha}{18\pi}\tc+ 2m^2g^{\mu\nu}  \ta -\frac{\alpha \mathscr{L}}{6\pi}- \frac{17\alpha}{72\pi}\tc & \DS, 
\end{cases}
\end{align} 
where we put the finite part of the counterterm of $g^{\mu\nu}$ in $V_1+V_2$.

The corresponding results for the renormalized form factors are
\begin{align}
A^R_e(0) & = 1 -\frac{2\alpha \mathscr{L}}{3\pi}-\frac{17}{18}\frac{\alpha}{\pi}, \label{AeMS}\\
A^R_\gamma(0) & = \frac{2\alpha \mathscr{L}}{3\pi}+\frac{17}{18}\frac{\alpha}{\pi},\\
\bar C^R_e(0) &= 
\begin{cases}
\frac{\alpha \mathscr{L}}{6\pi}-\frac{\alpha}{18\pi}, & \MS, \\
\frac{\alpha \mathscr{L}}{6\pi}+\frac{17}{72}\frac{\alpha}{\pi}, & \DS,
\end{cases} \\
\bar C^R_\gamma(0) &= 
\begin{cases}
-\frac{\alpha \mathscr{L}} {6\pi}+\frac{\alpha}{18\pi}, & \MS, \\
- \frac{\alpha \mathscr{L}}{6\pi}-\frac{17}{72}\frac{\alpha}{\pi}, & \DS. 
\end{cases}
\label{CgOS}
\end{align}

\section{Mass Sum Rules}
\label{sec_decompositions}
Different mass sum rules for the nucleon exist in the literature: a four-term decomposition proposed by Ji in Ref.~\cite{Ji:1995sv}, a two-term and a four-term decomposition by Lorc\'e~\cite{Lorce:2017xzd}, as well as a two-term decomposition of the mass squared by Hatta, Rajan, Tanaka~\cite{Hatta:2018sqd}. 
In the following, we will explore these sum rules for the electron, based on the one-loop results for the EMT discussed in the previous section.

\subsection{Two-term decompositions}
We start with the two-term decomposition of $m^2$ proposed in Ref.~\cite{Hatta:2018sqd} which reads
\begin{equation}
m^2 = \frac{1}{2}\ta \braket{\ta T_{e,R}\tc^\mu_\mu}  + \braket{\ta T_{\gamma,R}\tc^\mu_\mu}\tc \equiv \bar{m}_e^2 + \bar{m}_\gamma^2.\label{Hatta_traces|}
\end{equation}
From Eqs.~\eqref{trace_mixing-1}--\eqref{trace_mixing-2}, we find
\begin{align}
\frac{\bar{m}_e^2}{m^2} &= A_e^R(0) + 4\bar C_e^R(0) = 1+y-\gamma_m = \begin{cases}
1-\frac{7\alpha}{6\pi}, & \MS \\
 1, & \DS
\end{cases},\\
\frac{\bar{m}_{\gamma}^2}{m^2} &=A_\gamma^R(0) + 4\bar C_\gamma^R(0)= \gamma_m-y = \begin{cases}
\frac{7\alpha}{6\pi}, & \MS \\
0, & \DS 
\end{cases}
,
\end{align}
where we used $\braket{(m\bar\psi\psi)_R} = 2m^2(1-\gamma_m)$ and neglected  $O(\alpha^2)$ terms.
We observe that, at $O(\alpha)$, in the $\DS$ scheme the electron mass is exclusively related to the trace of the renormalized electron operator, while the photon contribution vanishes.
Once higher orders are taken into account one would find $\bar{m}_{\gamma}^2 \neq 0$ in the $\DS$ scheme.
However, to any order in perturbation theory the $\DS$ scheme ensures that $\bar{m}_e^2$ is exclusively given by a fermion operator and $\bar{m}_\gamma^2$ by a photon operator.
Therefore, the $\DS$ scheme is perhaps the most natural scheme for the two-term decomposition of $m^2$ proposed in Ref.~\cite{Hatta:2018sqd}.
We also point out that one can hardly assign a physical interpretation to both the size and the sign of the $O(\alpha)$ corrections, which (can) both depend on the scheme.
 
The two-term sum rule for $m^2$ of Ref.~\cite{Hatta:2018sqd} has the advantage of being a frame-independent decomposition. 
All the other decompositions that we consider in the following depend on the reference frame.  
We therefore discuss them first for the rest frame of the electron and afterwards comment on the required modifications in a moving frame.
 
In the two-term decomposition of Ref.~\cite{Lorce:2017xzd}, the mass of a particle is written as the sum of the energies carried by the constituent  and gauge degrees of freedom (electron and photon in our case),
\begin{equation}
m = U_e + U_\gamma.
\end{equation}
The definition of the (partial) energies $U_i$, in terms of renormalized operators$\,$\footnote{Operator renormalization was not discussed in any detail in Ref.~\cite{Lorce:2017xzd}},  is
\begin{align}
U_i &= \frac{\braket{\ta \int d^3\mathbf{x} \,T_i^{00}(0,\mathbf{x})  \tc_R}}{\braket{e(P)|e(P)}}\Bigg|_{\mathbf{P}=0} =  m\ta A^R_{i}(0) + \bar C^R_{i}(0)\tc,
\label{firstEnergies}
\end{align}
 where in the numerator we integrate over the volume to get an energy rather than an energy density.
(Inserted integration in numerator of above equation.)
We can therefore use the results in Eqs.~(\ref{AeMS})--(\ref{CgOS}) to compute the partial energies in the two renormalization schemes,
\begin{align}
U_e &= \begin{cases}
m\ta 1-\frac{\alpha \mathscr{L}}{2\pi}-\frac{\alpha}{\pi}\tc, & \MS\\
m\ta 1-\frac{\alpha \mathscr{L}}{2\pi}-\frac{17\alpha}{24\pi}\tc, & \DS
\end{cases}, \qquad 
U_\gamma = \begin{cases}
m\ta \frac{\alpha \mathscr{L}}{2\pi}+\frac{\alpha}{\pi}\tc, & \MS \\
m\ta \frac{\alpha \mathscr{L}}{2\pi}+\frac{17\alpha}{24\pi}\tc, & \DS
\end{cases}.
\end{align}
We find positive values for $U_\gamma$ in either scheme (unless the renormalization scale $\mu$ is extremely low), in agreement with what one would intuitively expect for the contribution due to the photon energy.
But we repeat that the interpretation of scheme-dependent renormalized operators has to be taken with care.
Below we will further comment on the properly renormalized operators associated with the two-term decomposition of Ref.~\cite{Lorce:2017xzd}.

\subsection{Four-term decompositions}
We would now like to comment on the four-term sum rule proposed in Ref.~\cite{Ji:1995sv} and studied for the electron for the first time in Ref.~\cite{Ji:1998bf}. 
In the latter paper, the individual contributions  to the mass decomposition are defined in terms of  the bare operators instead of the renormalized composite operators introduced in the previous section. 
Following Ref.~\cite{Ji:1995sv}, we can decompose the EMT into a trace part and a traceless part according to
\begin{equation}
T^{\mu\nu} = \hat T^{\mu\nu} + \bar T^{\mu\nu},
\end{equation}
with the trace term given by $\hat T^{\mu\nu}=\tfrac{1}{4}g^{\mu\nu}\,T^\alpha_\alpha$.
As described in the previous section, the separation of the two operators in terms of electron and photon contributions depends on the renormalization scheme and involves mixing of the electron and photon contribution under renormalization.
Therefore, the procedure of Ref.~\cite{Ji:1995sv}, where the traceless partial operators are obtained by subtracting the trace term from the full EMT separately for the electron and photon, deserves a fresh look. 
In accordance with Ref.~\cite{Ji:1995sv}, we introduce the QED Hamiltonian $H$ and the Hamiltonian density ${\cal H}$ as
\begin{equation}
H=\int d^3\mathbf{x} \,T^{00}(0,\mathbf{x})=\int d^3\mathbf{x} \,{\cal H}(0,\mathbf{x}).
\end{equation}
The separate electron and photon contributions to the traceless and trace operators are then defined as$\,$\footnote{The label indicates that here we are using  the definitions of  Ref.~\cite{Ji:1995sv}, which will be revised below. Note that
we use the covariant derivative defined as $D^i= \partial^i+ieA^i $, which differs by a global minus sign w.r.t. the definition of Ref.~\cite{Ji:1995sv}.}
\begin{align}
({\cal H}_e')_{\text{\cite{Ji:1995sv}}} &=\left[ (\bar T_e^{00})_{R}\right]_{\text{\cite{Ji:1995sv}}} = \ta \psi^\dagger \ta i\vet{D}\cdot\vet{\alpha}\tc\psi \tc_{R}+ \frac{3}{4}m\bar\psi \psi,\\
({\cal H}_m')_{\text{\cite{Ji:1995sv}}} &=\left [(\hat T_e^{00})_{R}\right]_{\text{\cite{Ji:1995sv}}} = \frac{1+\gamma_m}{4}m\bar\psi\psi,\\
({\cal H}_\gamma')_{\text{\cite{Ji:1995sv}}} &= \left[(\bar T_\gamma^{00})_{R}\right]_{\text{\cite{Ji:1995sv}}}= \frac{1}{2}\ta E^2 + B^2\tc_{R} ,\\
({\cal H}_a')_{\text{\cite{Ji:1995sv}}} &=\left[ (\hat T_\gamma^{00})_{R}\right]_{\text{\cite{Ji:1995sv}}}= -\frac{\beta(e)}{4e}\ta E^2 - B^2\tc_{R} .
\end{align}
Following Refs.~\cite{Ji:1995sv}, we can define
\begin{align}
\left({\cal H}_e\right)_{\text{\cite{Ji:1995sv}}} &\equiv[(\tilde{T}_e^{00})_R]_{\text{\cite{Lorce:2017xzd}}} = \left({\cal H}_e'\right)_{\text{\cite{Ji:1995sv}}}  + c_e \left({\cal H}'_m\right)_{\text{\cite{Ji:1995sv}}} , \label{He}\\
({\cal H}_m)_{\text{\cite{Ji:1995sv}}}  &\equiv[(\check{T}_e^{00})_R]_{\text{\cite{Lorce:2017xzd}}}= (1-c_e)({\cal H}_m')_{\text{\cite{Ji:1995sv}}} , \label{Hm} \\
({\cal H}_\gamma)_{\text{\cite{Ji:1995sv}}}  &\equiv[(\tilde{T}_\gamma^{00})_R]_{\text{\cite{Lorce:2017xzd}}}=  ({\cal H}_\gamma')_{\text{\cite{Ji:1995sv}}}  + c_\gamma ({\cal H}_a')_{\text{\cite{Ji:1995sv}}}, \label{Hgamma} \\
({\cal H}_a)_{\text{\cite{Ji:1995sv}}}  &\equiv[(\check{T}_\gamma^{00})_R]_{\text{\cite{Lorce:2017xzd}}} = (1-c_\gamma)({\cal H}_a' )_{\text{\cite{Ji:1995sv}}}\label{Ha},
\end{align}
where we also give reference to the corresponding nomenclature from  Ref.~\cite{Lorce:2017xzd} in terms of the $\tilde{T}_i^{00}$ and $\check{T}_i^{00}$ components of the EMT.
Choosing for the constants $c_i$ the values
\begin{equation}
c_e=\frac{-3}{1+\gamma_m}, \quad c_\gamma = 0, \label{OldContant}
\end{equation}
we then obtain the definitions of Ref.~\cite{Ji:1995sv}, 
\begin{align}
({\cal H}_e)_{\text{\cite{Ji:1995sv}}} & = \ta \psi^\dagger \ta i\vet{D}\cdot\vet{\alpha}\tc\psi \tc_{R}, \label{He_Ji} \\
({\cal H}_m)_{\text{\cite{Ji:1995sv}}} & =  \frac{4+\gamma_m}{4}m\bar\psi\psi, \\
({\cal H}_\gamma)_{\text{\cite{Ji:1995sv}}} &= ({\cal H}_\gamma')_{\text{\cite{Ji:1995sv}}}, \\
({\cal H}_a)_{\text{\cite{Ji:1995sv}}} &= ({\cal H}_a')_{\text{\cite{Ji:1995sv}}} \label{Ha_Ji}, 
\end{align}
where ${\cal H}_e$ represents the electron kinetic and potential energy, ${\cal H}_m$ is the quark mass contribution, ${\cal H}_\gamma$ is the photon kinetic and potential energy, and ${\cal H}_a$ is the anomaly contribution.
We can also introduce the two parameters $a$ and $b$ of Ref.~\cite{Ji:1995sv} as the matrix elements of the traceless and trace electron contributions, respectively,
\begin{equation}
\frac{3}{2}m^2 a_{\text{\cite{Ji:1995sv}}}  = \braket{(H_e')_{\text{\cite{Ji:1995sv}}} }_{\mathbf{P}=0}, \quad 2m^2 b_{\text{\cite{Ji:1995sv}}}  = \braket{(H_m')_{\text{\cite{Ji:1995sv}}} }_{\mathbf{P}=0}.
\end{equation}
Using the constraints in Eq.~\eqref{veryFundamentalSumRules}, we also obtain the relations
\begin{equation}
\frac{3}{2}m^2 (1-a_{\text{\cite{Ji:1995sv}}})  = \braket{(H_\gamma')_{\text{\cite{Ji:1995sv}}} }_{\mathbf{P}=0}, \quad 2m^2 (1-b_{\text{\cite{Ji:1995sv}}})  = \braket{(H_a')_{\text{\cite{Ji:1995sv}}} }_{\mathbf{P}=0}.
\end{equation}

So far we have reviewed the main points of the mass sum rule of Ref.~\cite{Ji:1995sv}.
In the following we address one issue of that paper and suggest a modification of the sum rule. 
In Ref.~\cite{Ji:1995sv}, the results for the traceless photon and electron contributions have been obtained by subtracting from the full EMT the trace part calculated with the use of the equations of motion for the fermionic fields. 
However, as already discussed in  Refs.~\cite{Tanaka:2018nae,Collins:1984xc},  this manipulation can not be applied when dealing with the renormalized operators $\mathcal{O}_i^R$, since the trace operation and the renormalization do not commute, i.e. $g_{\mu\nu}(F^{\mu\lambda}F^\nu_\lambda)_R\ne (F^{\mu\lambda} F_{\mu\lambda})_R$ and $g_{\mu\nu}(i\bar\psi\gamma^{(\mu}\overset{\leftrightarrow}{D}\ ^{\nu)}\psi)_R\ne(i\bar\psi\gamma^{(\lambda}\overset{\leftrightarrow}{D}\ _{\lambda)}\psi)_R$.
If instead we use the correct renormalized traceless electron and photon operators $\widetilde{\mathcal{O}}_1^R$ and $\widetilde{\mathcal{O}}_3^R$ in Eqs.~(\ref{traceless_photon_renorm})--(\ref{treceless_electron_renorm}), we find that the $00$-component of the traceless electron and photon parts are given by
\begin{align}
{\cal H}_e' &=(\bar T_e^{00})_R=\ta \psi^\dagger \ta i\vet{D}\cdot\vet{\alpha}\tc\psi \tc_R+ m\bar\psi \psi -\frac{1+y}{4}m\bar\psi\psi-\frac{x}{4}\ta F^{\mu\nu}F_{\mu\nu}\tc_R,
\label{Hprime-e-corretto}\\
{\cal H}_\gamma' &=(\bar T_\gamma^{00})_R= \frac{1}{2}\ta E^2 + B^2\tc_R + \frac{y-\gamma_m}{4}m\bar\psi\psi +\frac{1}{2} \ta \frac{\beta(e)}{2e}-x\tc \ta E^2 - B^2\tc_R.
\end{align}
The matrix element of the  revised expression in Eq.~\eqref{Hprime-e-corretto} for the $00$-component of the traceless electron operator allows us to identify the parameter $a$ with the renormalized form factor $A_e^R(0)$.

The $00$-components of the trace parts also change because of additional mixing, as can be seen from Eqs.~\eqref{trace_mixing-1}--\eqref{trace_mixing-2}. 
We find
\begin{align}
{\cal H}_m' &=(\hat T_e^{00})_R= \frac{1+y}{4}m\bar\psi\psi + \frac{x}{4}\ta F^{\mu\nu}F_{\mu\nu}\tc_R, \\
{\cal H}_a' &=(\hat T_\gamma^{00})_R= \frac{\gamma_m-y}{4}m\bar\psi\psi -\frac{1}{2} \ta \frac{\beta(e)}{2e}-x\tc\ta E^2 - B^2\tc_R.\label{Hprime-a-corretto}
\end{align}
To recover the intuitive picture in terms of kinetic and potential energy of the electron and photon, we need to take different combinations of the operators according to
\begin{align}
{\cal H}_e &\equiv[(\tilde{T}_e^{00})_R]= {\cal H}_e' + c_{em} {\cal H}'_m + c_{ea}{\cal H}'_a, \label{HeUs}\\
{\cal H}_m &\equiv [(\check{T}_e^{00})_R]=(1-c_{em}-c_{\gamma m}){\cal H}_m'+c_{ma}{\cal H}_a', \label{HmUs} \\
{\cal H}_\gamma &\equiv[(\tilde{T}_\gamma^{00})_R]= {\cal H}_\gamma' + c_{\gamma m} {\cal H}'_m + c_{\gamma a}{\cal H}'_a, \label{HgammaUs} \\
{\cal H}_a &\equiv[(\check{T}_\gamma^{00})_R]= (1-c_{ea}-c_{\gamma a}-c_{ma}){\cal H}_a', \label{HaUs}
\end{align}
with the constants
\begin{align}
c_{em} & = \frac{ (3-y) \frac{\beta(e)}{2e}- x (3-\gamma_m)}{- (1+ y)\frac{\beta(e)}{2e}   + x (1+\gamma_m)}, \label{NewConstant_1} \\
c_{ea} & = \frac{4 x}{(1 + y) \frac{\beta(e)}{2e} - x (1 + \gamma_m)}, \label{NewConstant_2} \\
c_{\gamma m}& = 0, \label{NewConstant_3} \\
c_{\gamma a} &= 1 \label{NewConstant_4} , \\
c_{ma} &= -c_{ea}.
\label{NewConstant_5}
\end{align}
This leads to the definitions
\begin{align}
{\cal H}_e &= \ta\psi^\dagger \ta i\vet{D}\cdot\vet{\alpha}\tc\psi\tc_R, \label{correctHe}\\
{\cal H}_m & =  m\bar\psi\psi,\\
{\cal H}_\gamma & = \frac{1}{2}\ta E^2 + B^2\tc_R, \label{correctHgamma} \\
{\cal H}_a &  = 0.\label{correctHa}
\end{align}
We argue that Eqs.~\eqref{correctHe}--\eqref{correctHa} are the appropriate operators for the mass sum rule if one follows the overall logic of Ji's original work, but uses the properly renormalized $00$-components of the traceless parts of the EMT for the fermion and the gauge field. 
Generally, the renormalized operators are no longer purely electron or photon operators (cf.~Eqs.~\eqref{renormOP1}-\eqref{renormOP4}).
It is also noteworthy that the expressions in Eqs.~\eqref{correctHe}--\eqref{correctHa} coincide formally with the classical results, i.e. the results one would obtain from the classical electromagnetic Lagrangian without the inclusion of the trace anomaly.
We have arrived at a decomposition with three nontrivial terms only.
Note that the vanishing of ${\cal H}_a$ is a general result and not limited to the one-loop perturbative treatment.
(Further discussion about the anomaly and its relation to the mass sum rule can be found in Ref.~\cite{Metz:2020vxd}.)
We emphasize that our analysis leading to Eqs.~\eqref{correctHe}--\eqref{correctHa} also holds for QCD.

We can also work out the revised expressions of the constants $a,b$, defined as the (correct) traceless and trace electron contributions, 
\begin{align}
\frac{3}{2}m^2a &= \braket{\widetilde{\mathcal{O}}_{3,R}^{00}}_{\mathbf{P}=0},\\
2m^2b &= \braket{(1+\gamma_m)m\bar\psi\psi}_{\mathbf{P}=0},\\
\frac{3}{2}m^2(1-a) &= \braket{\widetilde{\mathcal{O}}_{1,R}^{00}}_{\mathbf{P}=0},\\
2m^2(1-b) &= \frac{\beta(e)}{2e}\braket{\ta F^{\mu\nu}F_{\mu\nu}\tc_R}_{\mathbf{P}=0}.
\end{align}
We stress that $b$ is not directly the trace of the renormalized quark operator. 
Using the above definitions and Eqs.~(\ref{HeUs})-(\ref{HaUs}), we have the following mass decomposition:
\begin{align}
m_e & = \frac{3}{4}ma +\frac{m}{4}\ta x(1-b)\frac{2e}{\beta(e)} + b\frac{y-3}{1+\gamma_m}\tc ,  \label{new-mass1} \\
m_m &= \frac{mb}{1+\gamma_m}, \\
m_\gamma &= \frac{3}{4}m(1-a) + \frac{m(1-b)}{4}\ta 1-x\frac{2e}{\beta(e)}\tc + mb\frac{\gamma_m-y}{4(1+\gamma_m)} , \\
m_a &= 0,  \label{new-mass4}
\end{align}
where 
\begin{align}
m_i = \frac{\langle H_i\rangle}{\braket{e(P)|e(P)}}\Big|_{\mathbf{P}=0}.
\label{m_H}
\end{align}
In the two renormalization schemes,  the results at $O(\alpha)$ read
\begin{eqnarray} \label{Ji_mod_1loop_a}
&&\frac{m_e}{m} = \begin{cases}
\frac{\alpha}{2\pi} - \frac{\alpha \mathscr{L}}{2\pi}, & \MS \\
\frac{19\alpha}{24\pi} - \frac{\alpha \mathscr{L}}{2\pi}, & \DS 
\end{cases} ,\qquad\,
\frac{m_m}{m} = \begin{cases}
1- \frac{3\alpha}{2\pi}, & \MS \\
1- \frac{3\alpha}{2\pi}, & \DS 
\end{cases},
 \\
 \label{Ji_mod_1loop_b}
 &&
\frac{m_\gamma}{m} = \begin{cases}
\frac{\alpha}{\pi} + \frac{\alpha \mathscr{L}}{2\pi}, & \MS \\
\frac{17\alpha}{24\pi} + \frac{\alpha \mathscr{L}}{2\pi}, & \DS 
\end{cases},
\end{eqnarray}
Equipped with the proper one-loop results for the renormalized operators that appear in Eqs.~\eqref{correctHe}--\eqref{correctHa}, one can readily show that at one loop the terms in Eqs.~\eqref{He_Ji}--\eqref{Ha_Ji} do not add up to the mass of the electron.
This is just a consequence of the aforementioned issue with the sum rule in Ref.~\cite{Ji:1995sv}.

Before moving on to the second four-term sum rule, we make a brief comparison with the two-term decomposition of Ref.~\cite{Lorce:2017xzd}.
By means of Eqs.~\eqref{firstEnergies}, \eqref{Hprime-e-corretto}--\eqref{Hprime-a-corretto}, and~\eqref{correctHe}--\eqref{correctHa}, we find
\begin{equation}
U_e = m_e + m_m, \qquad U_\gamma = m_\gamma .
\label{2term_3term}
\end{equation}
Our three-term sum rule above could therefore be considered a refinement of the two-term decomposition of Ref.~\cite{Lorce:2017xzd}.
The relations in~\eqref{2term_3term} also allow one to readily identify the properly renormalized operators for $U_e$ and $U_\gamma$.

In Ref.~\cite{Lorce:2017xzd}, another type of four-term decomposition has been discussed, which makes use of the concept of energy introduced in Eq.~\eqref{firstEnergies} and of the partial pressure-volume work $W_i^j$ in the directions $j=x,y,z$, 
\begin{align}
W_i^j&= \frac{ \braket{\big( \int d^3\mathbf{x} \,T_i^{jj}(0,\mathbf{x})  \big)_R} }{\braket{e(P)|e(P)}}\Bigg|_{\mathbf{P}=0} .
\end{align}
While we follow here the general logic of Ref.~\cite{Lorce:2017xzd}, we (again) pay close attention to the operator renormalization. 
The partial energies and pressure-volume works can be related to the matrix elements of the operators $(\bar T^{00}_i)_R$ and $(\hat T^{00}_i)_R$ according to
\begin{equation}
\frac{ \braket{\big( \int d^3\mathbf{x} \, \bar{T}_i^{00}(0,\mathbf{x})  \big)_R}}{\braket{e(P)|e(P)}}\Bigg|_{\mathbf{P}=0} = \frac{3}{4}(U_i + W_i), 
\quad \frac{ \braket{\big( \int d^3\mathbf{x} \, \hat{T}_i^{00}(0,\mathbf{x})  \big)_R}}{\braket{e(P)|e(P)}}\Bigg|_{\mathbf{P}=0} = \frac{1}{4}(U_i -3 W_i),
\label{StartingPointLorce}
\end{equation}
where $3W_i=W_i^{x}+W_i^{y}+W_i^{z}$.
The four term decomposition of Ref~\cite{Lorce:2017xzd} reads as
\begin{equation}
m = \tilde U_e + \tilde U_\gamma + \check U_e + \check U_\gamma,\label{new-dec-cedric}
\end{equation}
where the individual terms correspond to the  contributions of the internal energy to the matrix elements of the $\hat{T}_i^{00}$ and $\check{T}^{00}_i$ operators defined in Eqs.~\eqref{He}-\eqref{Ha}.
Using the properly renormalized operators in Eqs.~(\ref{HeUs})--(\ref{HaUs}), we obtain:
\begin{align}
\tilde U_e & = \frac{U_e}{4}\ta 3 + c_{em}\tc +  \frac{U_\gamma}{4}c_{ea}, & \check U_e & = \frac{U_e}{4}\ta  1-c_{em}-c_{\gamma m}\tc + \frac{U_\gamma}{4}c_{ma},\\
\tilde U_\gamma & = \frac{U_\gamma}{4}\ta 3 + c_{\gamma a}\tc +  \frac{U_e}{4}c_{\gamma m}, & \check U_\gamma & = \frac{U_\gamma}{4}\ta 1-c_{ea}-c_{\gamma a}-c_{ma}\tc,
\end{align}
with the constants $c_{i}$ defined in Eqs.~\eqref{NewConstant_1}--\eqref{NewConstant_5}.
The main difference with respect to Ref.~\cite{Lorce:2017xzd} is that we need to mix $(\bar T^{00}_e)_R$ with $(\hat T^{00}_e)_R$ and $(\hat T^{00}_\gamma)_R$,  $(\bar T^{00}_\gamma)_R$ with $(\hat T^{00}_\gamma)_R$ and $(\hat T^{00}_e)_R$ with $(\hat T^{00}_\gamma)_R$.
Using the coefficients in Eqs.~\eqref{NewConstant_3}--\eqref{NewConstant_5} we find for the photon sector
\begin{equation}
\tilde{U}_\gamma = U_\gamma, \qquad \check{U}_\gamma = 0 .
\end{equation}
This means that, once working with properly renormalized operators, the four-term sum rule of Ref.~\cite{Lorce:2017xzd} in fact reduce only to three nontrivial contributions.
Finally, in the two renormalization schemes we have the explicit results
\begin{eqnarray}
\frac{\tilde U_e}{m} &=& \begin{cases}
\frac{\alpha}{3\pi}, & \MS \\
\frac{3\alpha}{2\pi}, & \DS 
\end{cases},
\qquad\qquad\qquad\,\,\,
\frac{\check U_e }{m} =\begin{cases}
1- \frac{4\alpha}{3\pi} - \frac{\alpha \mathscr{L}}{2\pi},& \MS \\
1- \frac{53\alpha}{24\pi} - \frac{\alpha \mathscr{L}}{2\pi}, & \DS 
\end{cases},
\\
\frac{\tilde U_\gamma}{m}  &=& \begin{cases}
\frac{\alpha}{\pi}  +\frac{\alpha \mathscr{L}}{2\pi},& \MS \\
\frac{17\alpha}{24\pi} + \frac{\alpha \mathscr{L}}{2\pi}, & \DS 
\end{cases}. 
\end{eqnarray}

\subsection{Sum rules in a moving frame}
Except the two-term decomposition of Ref.~\cite{Hatta:2018sqd}, all other mass sum rules, strictly speaking, only hold in the rest frame.
However, one may expect that in a moving frame they still provide a meaningful result.
In fact they become energy decompositions as we discuss in the following (see Ref.~\cite{Lorce:2018egm} for a more general discussion on the frame dependence of the matrix elements of the EMT).
For a moving electron with energy $E$, the partial energies become
\begin{equation}
U_i = EA_i^R(0) + \frac{m^2}{E}\bar C_i^R(0).
\end{equation}
If the electron momentum points along the $\hat z$ axis, i.e. $P^\mu = (E,0,0,p)$, we find for the partial pressure-volume works
\begin{equation}
W_i^{x} =W_i^y= -\frac{m^2}{E}\bar C_i^R(0), \quad W_i^z = \frac{E^2-m^2}{E}A_i^R(0)-\frac{m^2}{E}\bar C_i^R(0),
\end{equation}
and therefore
\begin{equation}
W_i= \frac{E^2-m^2}{3E}A_i^R(0)-\frac{m^2}{E}\bar C_i^R(0).
\end{equation}
The values of the $a,b$ coefficients are not modified in a moving frame since they are related to the form factors and not to the energy.
Recalling the identification $a = A_e^R(0)$, we obtain the following modification of the expectation values of the traceless operators:
\begin{align}
\frac{\braket{\int d^3\mathbf{x} \, \widetilde{\mathcal{O}}_{3,R}^{00}}}{\braket{e(P)|e(P)}}\Big|_{\mathbf{P}=0} &= \frac{3}{4}am \rightarrow a\ta E - \frac{m^2}{4E}\tc ,\\
\frac{\braket{\int d^3\mathbf{x} \, \widetilde{\mathcal{O}}_{1,R}^{00}}}{\braket{e(P)|e(P)}}\Big|_{\mathbf{P}=0} &= \frac{3}{4}(1-a)m \rightarrow (1-a)\ta E - \frac{m^2}{4E}\tc.
\end{align}
The trace parts are affected too because of the normalization of the states.  
We have
\begin{align}
\frac{\braket{\int d^3\mathbf{x} \, (1+\gamma_m)m\bar\psi\psi}}{\braket{e(P)|e(P)}}\Big|_{\mathbf{P}=0}& = mb \rightarrow b\frac{m^2}{E},\\
\frac{\beta(e)}{2e}\frac{\braket{ \int d^3\mathbf{x}  \ta F^{\mu\nu}F_{\mu\nu}\tc_R}}{\braket{e(P)|e(P)}}\Big|_{\mathbf{P}=0} &= m(1-b) \rightarrow \frac{m^2}{E}(1-b).
\end{align}
These results allow us to obtain the counterparts of Eqs.~\eqref{Ji_mod_1loop_a}--\eqref{Ji_mod_1loop_b} for a moving frame,
\begin{eqnarray}
&&\frac{m_e}{E} = \begin{cases}
\frac{E^2-m^2}{E^2} + \frac{\alpha}{\pi}\ta -\frac{17}{18} +\frac{13m^2}{9E^2}-\frac{2 \mathscr{L} }{3} + \frac{\mathscr{L}m^2}{6E^2} \tc, & \MS \\
\frac{E^2-m^2}{E^2} + \frac{\alpha}{\pi}\ta -\frac{17}{18}+\frac{125m^2}{72E^2} - \frac{2 \mathscr{L} }{3} + \frac{\mathscr{L}m^2}{6E^2} \tc,& \DS 
\end{cases} ,
\quad
\frac{m_m}{E} = \begin{cases}
\frac{m^2}{E^2} \ta 1- \frac{3\alpha}{2\pi}\tc , & \MS \\
\frac{m^2}{E^2}\ta 1- \frac{3\alpha}{2\pi}\tc, & \DS 
\end{cases},\nonumber
\\
&&\frac{m_\gamma}{E} = \begin{cases}
\frac{\alpha}{\pi} \ta \frac{17}{18} +\frac{m^2}{18E^2} +\frac{2 \mathscr{L} }{3} - \frac{\mathscr{L}m^2}{6E^2}\tc,& \MS \\
\frac{\alpha}{\pi}\ta \frac{17}{18}  - \frac{17m^2}{72E^2}+ \frac{2 \mathscr{L} }{3} - \frac{ \mathscr{L}m^2}{6E^2} \tc,& \DS 
\end{cases} ,
\label{Ji_mod_moving}
\end{eqnarray}
while for the decomposition in Eq.~\eqref{new-dec-cedric} we obtain
\begin{eqnarray}
&&
\frac{\tilde U_e}{E} = \begin{cases}
\frac{\alpha}{3\pi},& \MS  \\
\frac{3\alpha}{2\pi},& \DS 
\end{cases} ,\quad\qquad\quad\qquad\quad\quad
\frac{\check U_e}{E}= \begin{cases}
1 + \frac{\alpha}{\pi}\ta -\frac{23}{18} - \frac{m^2}{18E^2} - \frac{2 \mathscr{L}}{3} + \frac{\mathscr{L} m^2}{6E^2} \tc,& \MS \\
1 + \frac{\alpha}{\pi}\ta -\frac{22}{9} + \frac{17m^2}{72E^2} - \frac{2 \mathscr{L} }{3} + \frac{\mathscr{L}m^2 }{6E^2} \tc,& \DS 
\end{cases},\nonumber
 \\
 &&
\frac{\tilde U_\gamma}{E}  =\begin{cases}
\frac{\alpha}{\pi}\ta \frac{17}{18} + \frac{m^2}{18E^2} + \frac{2\mathscr{L}}{3} - \frac{\mathscr{L}m^2}{6E^2} \tc,& \MS \\
\frac{\alpha}{\pi}\ta \frac{17}{18} - \frac{17m^2}{72E^2} + \frac{2\mathscr{L}}{3} - \frac{\mathscr{L}m^2}{6E^2} \tc,& \DS 
\end{cases} .
\label{Lorce_mod_moving}
\end{eqnarray}
One can readily verify that the terms in Eq.~\eqref{Ji_mod_moving} and in Eq.~\eqref{Lorce_mod_moving} add up to $E$.
On the other hand, the individual terms of the energy decompositions cannot be obtained by multiplying the corresponding expressions in the rest frame by a common overall kinematic factor.

\section{Conclusions}
\label{sec_conclusions}
We discussed in detail the forward matrix elements of the EMT for an electron state by performing the calculation at order $O(\alpha)$ in QED.
In particular, we presented an explicit calculation of the EMT renormalization procedure described in Refs.~\cite{Hatta:2018sqd,Tanaka:2018nae}.   
We reviewed the mass sum rules proposed by Ji~\cite{Ji:1994av}, Lorc\'e~\cite{Lorce:2017xzd}, and Hatta, Rajan, Tanaka~\cite{Hatta:2018sqd} for the case of the nucleon, and applied them to the case of the electron by paying attention to the mixing of the individual contributions under renormalization. 
We also emphasized the scheme dependence of the various contributions to the electron mass which complicates the interpretation of the results.

In relation to the aforementioned papers on the nucleon mass our main findings are essentially threefold:
First, we propose a new renormalization scheme which is arguably the most natural one for the two-term decomposition of the squared mass $m^2 = \bar{m}_e^2 + \bar{m}_\gamma^2$ in Ref.~\cite{Hatta:2018sqd}.
In this scheme, $\bar{m}_e^2$ is exclusively given by a (renormalized) fermion operator and $\bar{m}_\gamma^2$ by a (renormalized) photon operator.
Second, we point at a nontrivial issue in the derivation of the four-term decomposition of~\cite{Ji:1994av}, which can be traced back to finding the properly renormalized operators for the trace of the EMT.
Once this point is corrected, one actually arrives at a decomposition that contains three terms only. 
Third, we identify renormalized operators for the two-term and four-term decompositions of Ref.~\cite{Lorce:2017xzd}. 
As a consequence, the aforementioned three-term decomposition (obtained in the spirit of Ref.~\cite{Ji:1994av}) can be considered a refinement of the two-term decomposition of Ref.~\cite{Lorce:2017xzd}, and the four-term decompositions of Ref.~\cite{Lorce:2017xzd} boils down to a three-term decomposition.

The present work suggests related future studies:~The implications of the findings for the various mass sum rules should be studied for the phenomenology of the nucleon mass.
Moreover, the one-loop QED calculation can be extended to the off-forward matrix elements of the EMT for the electron, which give access to pressure and shear distributions.
Work along those lines is in progress.

\begin{acknowledgments}
The work of A.M.~has been supported by the National Science Foundation under grant number PHY-1812359, and by the U.S. Department of Energy, Office of Science, Office of Nuclear Physics, within the framework of the TMD Topical Collaboration.
The work of B.P.~and S.R.~has been supported by the European Union's Horizon 2020 programme under grant agreement No.~824093(STRONG2020) and under the European Research Council (ERC) grant agreement No.~647981 (3DSPIN).
B.P.~and S.R.~are grateful to M.~Moretti for useful discussions.
The authors are also thankful to C. Lorc\'e for a careful reading of the manuscript.
S.R.~acknowledges the hospitality at the Department of Physics of Temple University (Philadelphia) where part of this work was performed.
\end{acknowledgments}

\begin{thebibliography}{10}

\bibitem{Diehl:2015uka}
M.~Diehl, \emph{{Introduction to GPDs and TMDs}},
  \href{https://doi.org/10.1140/epja/i2016-16149-3}{\emph{Eur. Phys. J.}
  {\bfseries A52} (2016) 149}
  [\href{https://arxiv.org/abs/1512.01328}{{\ttfamily 1512.01328}}].

\bibitem{Bacchetta:2016ccz}
A.~Bacchetta, \emph{{Where do we stand with a 3-D picture of the proton?}},
  \href{https://doi.org/10.1140/epja/i2016-16163-5}{\emph{Eur. Phys. J.}
  {\bfseries A52} (2016) 163}.

\bibitem{Belitsky:2003nz}
A.V.~Belitsky, X.-d.~Ji and F.~Yuan, \emph{{Quark imaging in the proton via
  quantum phase space distributions}},
  \href{https://doi.org/10.1103/PhysRevD.69.074014}{\emph{Phys. Rev.}
  {\bfseries D69} (2004) 074014}
  [\href{https://arxiv.org/abs/hep-ph/0307383}{{\ttfamily hep-ph/0307383}}].

\bibitem{Meissner:2009ww}
S.~Meissner, A.~Metz and M.~Schlegel, \emph{{Generalized parton correlation
  functions for a spin-1/2 hadron}},
  \href{https://doi.org/10.1088/1126-6708/2009/08/056}{\emph{JHEP} {\bfseries
  08} (2009) 056} [\href{https://arxiv.org/abs/0906.5323}{{\ttfamily
  0906.5323}}].

\bibitem{Lorce:2011dv}
C.~Lorc\'e, B.~Pasquini and M.~Vanderhaeghen, \emph{{Unified framework for
  generalized and transverse-momentum dependent parton distributions within a
  3Q light-cone picture of the nucleon}},
  \href{https://doi.org/10.1007/JHEP05(2011)041}{\emph{JHEP} {\bfseries 05}
  (2011) 041} [\href{https://arxiv.org/abs/1102.4704}{{\ttfamily 1102.4704}}].

\bibitem{Ji:1996ek}
X.-D.~Ji, \emph{{Gauge-Invariant Decomposition of Nucleon Spin}},
  \href{https://doi.org/10.1103/PhysRevLett.78.610}{\emph{Phys. Rev. Lett.}
  {\bfseries 78} (1997) 610}
  [\href{https://arxiv.org/abs/hep-ph/9603249}{{\ttfamily hep-ph/9603249}}].

\bibitem{Polyakov:2002yz}
M.V.~Polyakov, \emph{{Generalized parton distributions and strong forces inside
  nucleons and nuclei}},
  \href{https://doi.org/10.1016/S0370-2693(03)00036-4}{\emph{Phys. Lett.}
  {\bfseries B555} (2003) 57}
  [\href{https://arxiv.org/abs/hep-ph/0210165}{{\ttfamily hep-ph/0210165}}].

\bibitem{Polyakov:2018zvc}
M.V.~Polyakov and P.~Schweitzer, \emph{{Forces inside hadrons: pressure,
  surface tension, mechanical radius, and all that}},
  \href{https://doi.org/10.1142/S0217751X18300259}{\emph{Int. J. Mod. Phys.}
  {\bfseries A33} (2018) 1830025}
  [\href{https://arxiv.org/abs/1805.06596}{{\ttfamily 1805.06596}}].

\bibitem{Lorce:2015lna}
C.~Lorc\'e, \emph{{The light-front gauge-invariant energy-momentum tensor}},
  \href{https://doi.org/10.1007/JHEP08(2015)045}{\emph{JHEP} {\bfseries 08}
  (2015) 045} [\href{https://arxiv.org/abs/1502.06656}{{\ttfamily
  1502.06656}}].

\bibitem{Lorce:2018egm}
C.~Lorc\'e, H.~Moutarde and A.P.~Trawi\'nski, \emph{{Revisiting the mechanical
  properties of the nucleon}},
  \href{https://doi.org/10.1140/epjc/s10052-019-6572-3}{\emph{Eur.\ Phys.\ J.\
  C} {\bfseries 79} (2019) 89}
  [\href{https://arxiv.org/abs/1810.09837}{{\ttfamily 1810.09837}}].

\bibitem{Burkert:2018bqq}
V.D.~Burkert, L.~Elouadrhiri and F.X.~Girod, \emph{{The pressure distribution
  inside the proton}},
  \href{https://doi.org/10.1038/s41586-018-0060-z}{\emph{Nature} {\bfseries
  557} (2018) 396}.

\bibitem{Kumericki:2019ddg}
K.~Kumeri\v{c}ki, \emph{{Measurability of pressure inside the proton}},
  \href{https://doi.org/10.1038/s41586-019-1211-6}{\emph{Nature} {\bfseries
  570} (2019) E1}.

\bibitem{Hagler:2003jd}
{\scshape LHPC, SESAM} collaboration, \emph{{Moments of nucleon generalized
  parton distributions in lattice QCD}},
  \href{https://doi.org/10.1103/PhysRevD.68.034505}{\emph{Phys. Rev.}
  {\bfseries D68} (2003) 034505}
  [\href{https://arxiv.org/abs/hep-lat/0304018}{{\ttfamily hep-lat/0304018}}].

\bibitem{Gockeler:2003jfa}
{\scshape QCDSF} collaboration, \emph{{Generalized parton distributions from
  lattice QCD}},
  \href{https://doi.org/10.1103/PhysRevLett.92.042002}{\emph{Phys. Rev. Lett.}
  {\bfseries 92} (2004) 042002}
  [\href{https://arxiv.org/abs/hep-ph/0304249}{{\ttfamily hep-ph/0304249}}].

\bibitem{Hagler:2007xi}
{\scshape LHPC} collaboration, \emph{{Nucleon Generalized Parton Distributions
  from Full Lattice QCD}},
  \href{https://doi.org/10.1103/PhysRevD.77.094502}{\emph{Phys. Rev.}
  {\bfseries D77} (2008) 094502}
  [\href{https://arxiv.org/abs/0705.4295}{{\ttfamily 0705.4295}}].

\bibitem{Yang:2018nqn}
Y.-B.~Yang, J.~Liang, Y.-J.~Bi, Y.~Chen, T.~Draper, K.-F.~Liu et~al.,
  \emph{{Proton Mass Decomposition from the QCD Energy Momentum Tensor}},
  \href{https://doi.org/10.1103/PhysRevLett.121.212001}{\emph{Phys. Rev. Lett.}
  {\bfseries 121} (2018) 212001}
  [\href{https://arxiv.org/abs/1808.08677}{{\ttfamily 1808.08677}}].

\bibitem{Yang:2014xsa}
Y.-B.~Yang, Y.~Chen, T.~Draper, M.~Gong, K.-F.~Liu, Z.~Liu et~al., \emph{{Meson
  Mass Decomposition from Lattice QCD}},
  \href{https://doi.org/10.1103/PhysRevD.91.074516}{\emph{Phys. Rev.}
  {\bfseries D91} (2015) 074516}
  [\href{https://arxiv.org/abs/1405.4440}{{\ttfamily 1405.4440}}].

\bibitem{Shanahan:2018nnv}
P.E.~Shanahan and W.~Detmold, \emph{{Pressure Distribution and Shear Forces
  inside the Proton}},
  \href{https://doi.org/10.1103/PhysRevLett.122.072003}{\emph{Phys. Rev. Lett.}
  {\bfseries 122} (2019) 072003}
  [\href{https://arxiv.org/abs/1810.07589}{{\ttfamily 1810.07589}}].

\bibitem{Leader:2013jra}
E.~Leader and C.~Lorc\'e, \emph{{The angular momentum controversy: What's it
  all about and does it matter?}},
  \href{https://doi.org/10.1016/j.physrep.2014.02.010}{\emph{Phys. Rept.}
  {\bfseries 541} (2014) 163}
  [\href{https://arxiv.org/abs/1309.4235}{{\ttfamily 1309.4235}}].

\bibitem{Lorce:2017wkb}
C.~Lorc\'e, L.~Mantovani and B.~Pasquini, \emph{{Spatial distribution of
  angular momentum inside the nucleon}},
  \href{https://doi.org/10.1016/j.physletb.2017.11.018}{\emph{Phys. Lett.}
  {\bfseries B776} (2018) 38}
  [\href{https://arxiv.org/abs/1704.08557}{{\ttfamily 1704.08557}}].

\bibitem{Ji:1994av}
X.-D.~Ji, \emph{{A QCD analysis of the mass structure of the nucleon}},
  \href{https://doi.org/10.1103/PhysRevLett.74.1071}{\emph{Phys. Rev. Lett.}
  {\bfseries 74} (1995) 1071}
  [\href{https://arxiv.org/abs/hep-ph/9410274}{{\ttfamily hep-ph/9410274}}].

\bibitem{Ji:1995sv}
X.-D.~Ji, \emph{{Breakup of hadron masses and energy-momentum tensor of QCD}},
  \href{https://doi.org/10.1103/PhysRevD.52.271}{\emph{Phys. Rev.} {\bfseries
  D52} (1995) 271} [\href{https://arxiv.org/abs/hep-ph/9502213}{{\ttfamily
  hep-ph/9502213}}].

\bibitem{Roberts:2016vyn}
C.D.~Roberts, \emph{{Perspective on the origin of hadron masses}},
  \href{https://doi.org/10.1007/s00601-016-1168-z}{\emph{Few Body Syst.}
  {\bfseries 58} (2017) 5} [\href{https://arxiv.org/abs/1606.03909}{{\ttfamily
  1606.03909}}].

\bibitem{Lorce:2017xzd}
C.~Lorc\'e, \emph{{On the hadron mass decomposition}},
  \href{https://doi.org/10.1140/epjc/s10052-018-5561-2}{\emph{Eur. Phys. J.}
  {\bfseries C78} (2018) 120}
  [\href{https://arxiv.org/abs/1706.05853}{{\ttfamily 1706.05853}}].

\bibitem{Hatta:2018sqd}
Y.~Hatta, A.~Rajan and K.~Tanaka, \emph{{Quark and gluon contributions to the
  QCD trace anomaly}},
  \href{https://doi.org/10.1007/JHEP12(2018)008}{\emph{JHEP} {\bfseries 12}
  (2018) 008} [\href{https://arxiv.org/abs/1810.05116}{{\ttfamily
  1810.05116}}].

\bibitem{Kharzeev:1995ij}
D.~Kharzeev, \emph{{Quarkonium interactions in QCD}},
  \href{https://doi.org/10.3254/978-1-61499-215-8-105}{\emph{Proc. Int. Sch.
  Phys. Fermi} {\bfseries 130} (1996) 105}
  [\href{https://arxiv.org/abs/nucl-th/9601029}{{\ttfamily nucl-th/9601029}}].

\bibitem{Joosten:2018gyo}
S.~Joosten and Z.E.~Meziani, \emph{{Heavy Quarkonium Production at Threshold:
  from JLab to EIC}}, \href{https://doi.org/10.22323/1.308.0017}{\emph{PoS}
  {\bfseries QCDEV2017} (2018) 017}
  [\href{https://arxiv.org/abs/1802.02616}{{\ttfamily 1802.02616}}].

\bibitem{Hatta:2018ina}
Y.~Hatta and D.-L.~Yang, \emph{{Holographic $J/\psi$ production near threshold
  and the proton mass problem}},
  \href{https://doi.org/10.1103/PhysRevD.98.074003}{\emph{Phys. Rev.}
  {\bfseries D98} (2018) 074003}
  [\href{https://arxiv.org/abs/1808.02163}{{\ttfamily 1808.02163}}].

\bibitem{Ali:2019lzf}
{\scshape GlueX} collaboration, \emph{{First Measurement of Near-Threshold
  $J/\Psi$ Exclusive Photoproduction off the Proton}},
  \href{https://doi.org/10.1103/PhysRevLett.123.072001}{\emph{Phys. Rev. Lett.}
  {\bfseries 123} (2019) 072001}
  [\href{https://arxiv.org/abs/1905.10811}{{\ttfamily 1905.10811}}].

\bibitem{Hatta:2019lxo}
Y.~Hatta, A.~Rajan and D.-L.~Yang, \emph{{Near threshold $J/\psi$ and
  $\Upsilon$ photoproduction at JLab and RHIC}},
  \href{https://doi.org/10.1103/PhysRevD.100.014032}{\emph{Phys. Rev.}
  {\bfseries D100} (2019) 014032}
  [\href{https://arxiv.org/abs/1906.00894}{{\ttfamily 1906.00894}}].

\bibitem{Mamo:2019mka}
K.A.~Mamo and I.~Zahed, \emph{{Diffractive photoproduction of $J/\psi$ and
  $\Upsilon$ using holographic QCD: gravitational form factors and GPD of
  gluons in the proton}},
  \href{https://doi.org/10.1103/PhysRevD.101.086003}{\emph{Phys. Rev. D}
  {\bfseries 101} (2020) 086003}
  [\href{https://arxiv.org/abs/1910.04707}{{\ttfamily 1910.04707}}].

\bibitem{Wang:2019mza}
R.~Wang, J.~Evslin and X.~Chen, \emph{{The origin of proton mass from J/${\Psi
  }$ photo-production data}},
  \href{https://doi.org/10.1140/epjc/s10052-020-8057-9}{\emph{Eur. Phys. J. C}
  {\bfseries 80} (2020) 507}
  [\href{https://arxiv.org/abs/1912.12040}{{\ttfamily 1912.12040}}].

\bibitem{Bacchetta:2015qka}
A.~Bacchetta, L.~Mantovani and B.~Pasquini, \emph{{Electron in
  three-dimensional momentum space}},
  \href{https://doi.org/10.1103/PhysRevD.93.013005}{\emph{Phys.\ Rev.\ D}
  {\bfseries 93} (2016) 013005}
  [\href{https://arxiv.org/abs/1508.06964}{{\ttfamily 1508.06964}}].

\bibitem{Miller:2014vla}
G.A.~Miller, \emph{{Electron structure: Shape, size, and generalized parton
  distributions in QED}},
  \href{https://doi.org/10.1103/PhysRevD.90.113001}{\emph{Phys.\ Rev.\ D}
  {\bfseries 90} (2014) 113001}
  [\href{https://arxiv.org/abs/1409.7412}{{\ttfamily 1409.7412}}].

\bibitem{Hoyer:2009sg}
P.~Hoyer and S.~Kurki, \emph{{The Transverse shape of the electron}},
  \href{https://doi.org/10.1103/PhysRevD.81.013002}{\emph{Phys.\ Rev.\ D}
  {\bfseries 81} (2010) 013002}
  [\href{https://arxiv.org/abs/0911.3011}{{\ttfamily 0911.3011}}].

\bibitem{Brodsky:2000ii}
S.J.~Brodsky, D.S.~Hwang, B.-Q.~Ma and I.~Schmidt, \emph{{Light-cone
  representation of the spin and orbital angular momentum of relativistic
  composite systems}},
  \href{https://doi.org/10.1016/S0550-3213(00)00626-X}{\emph{Nucl. Phys.}
  {\bfseries B593} (2001) 311}
  [\href{https://arxiv.org/abs/hep-th/0003082}{{\ttfamily hep-th/0003082}}].

\bibitem{Berends:1975ah}
F.A.~Berends and R.~Gastmans, \emph{{Quantum Electrodynamical Corrections to
  Graviton-Matter Vertices}},
  \href{https://doi.org/10.1016/0003-4916(76)90245-1}{\emph{Annals Phys.}
  {\bfseries 98} (1976) 225}.

\bibitem{Milton:1976jr}
K.A.~Milton, \emph{{Quantum Electrodynamic Corrections to the Gravitational
  Interaction of the electron}},
  \href{https://doi.org/10.1103/PhysRevD.15.538}{\emph{Phys. Rev.} {\bfseries
  D15} (1977) 538}.

\bibitem{Milton:1977je}
K.A.~Milton, \emph{{Quantum Electrodynamic Corrections to the Gravitational
  Interaction of the Photon}},
  \href{https://doi.org/10.1103/PhysRevD.15.2149}{\emph{Phys. Rev.} {\bfseries
  D15} (1977) 2149}.

\bibitem{Ji:1998bf}
X.-D.~Ji and W.~Lu, \emph{{A Modern anatomy of electron mass}},
  \href{https://arxiv.org/abs/hep-ph/9802437}{{\ttfamily hep-ph/9802437}}.

\bibitem{Harindranath:1998ve}
A.~Harindranath and R.~Kundu, \emph{{On Orbital angular momentum in deep
  inelastic scattering}},
  \href{https://doi.org/10.1103/PhysRevD.59.116013}{\emph{Phys. Rev.}
  {\bfseries D59} (1999) 116013}
  [\href{https://arxiv.org/abs/hep-ph/9802406}{{\ttfamily hep-ph/9802406}}].

\bibitem{Burkardt:2008ua}
M.~Burkardt and H.~BC, \emph{{Angular Momentum Decomposition for an Electron}},
  \href{https://doi.org/10.1103/PhysRevD.79.071501}{\emph{Phys. Rev.}
  {\bfseries D79} (2009) 071501}
  [\href{https://arxiv.org/abs/0812.1605}{{\ttfamily 0812.1605}}].

\bibitem{Kanazawa:2014nha}
K.~Kanazawa, C.~Lorc\'e, A.~Metz, B.~Pasquini and M.~Schlegel, \emph{{Twist-2
  generalized transverse-momentum dependent parton distributions and the
  spin/orbital structure of the nucleon}},
  \href{https://doi.org/10.1103/PhysRevD.90.014028}{\emph{Phys. Rev.}
  {\bfseries D90} (2014) 014028}
  [\href{https://arxiv.org/abs/1403.5226}{{\ttfamily 1403.5226}}].

\bibitem{Liu:2014fxa}
T.~Liu and B.-Q.~Ma, \emph{{Angular momentum decomposition from a QED
  example}}, \href{https://doi.org/10.1103/PhysRevD.91.017501}{\emph{Phys.
  Rev.} {\bfseries D91} (2015) 017501}
  [\href{https://arxiv.org/abs/1412.7775}{{\ttfamily 1412.7775}}].

\bibitem{Ji:2015sio}
X.~Ji, A.~Sch{\"a}fer, F.~Yuan, J.-H.~Zhang and Y.~Zhao, \emph{{Spin
  decomposition of the electron in QED}},
  \href{https://doi.org/10.1103/PhysRevD.93.054013}{\emph{Phys. Rev.}
  {\bfseries D93} (2016) 054013}
  [\href{https://arxiv.org/abs/1511.08817}{{\ttfamily 1511.08817}}].

\bibitem{Belinfante1}
F.~Belinfante, \emph{{On the spin angular momentum of mesons}}, {\emph{Physica}
  {\bfseries 6} (1939) 887}.

\bibitem{Belinfante2}
F.~Belinfante, \emph{{On the current and the density of the electric charge,
  the energy, the linear momentum and the angular momentum of arbitrary
  fields}}, {\emph{Physica} {\bfseries 7} (1940) 449}.

\bibitem{Rosenfeld}
L.~Rosenfeld, \emph{{On the energy-momentum tensor}}, {\emph{M\'em. Acad. Roy.
  Belg.} {\bfseries 18} (1940) 1}.

\bibitem{Montesinos:2006th}
M.~Montesinos and E.~Flores, \emph{{Symmetric energy-momentum tensor in
  Maxwell, Yang-Mills, and proca theories obtained using only Noether's
  theorem}}, {\emph{Rev. Mex. Fis.} {\bfseries 52} (2006) 29}
  [\href{https://arxiv.org/abs/hep-th/0602190}{{\ttfamily hep-th/0602190}}].

\bibitem{Eriksen:1979vq}
E.~Eriksen and J.M.~Leinaas, \emph{{Gauge invariance and the transformation
  properties of the electromagnetic four potential}},
  \href{https://doi.org/10.1088/0031-8949/22/3/003}{\emph{Phys. Scripta}
  {\bfseries 22} (1980) 199}.

\bibitem{Takahashi:1985dt}
Y.~Takahashi, \emph{{Energy-momentum tensors in relativistic and
  nonrelativistic classical field theory}}, {\emph{Fortsch. Phys.} {\bfseries
  34} (1986) 323}.

\bibitem{Munoz:1996wp}
G.~Munoz, \emph{{Lagrangian field theories and energy-momentum tensors}},
  \href{https://doi.org/10.1119/1.18336}{\emph{Am. J. Phys.} {\bfseries 64}
  (1996) 1153}.

\bibitem{Tanaka:2018nae}
K.~Tanaka, \emph{{Three-loop formula for quark and gluon contributions to the
  QCD trace anomaly}},
  \href{https://doi.org/10.1007/JHEP01(2019)120}{\emph{JHEP} {\bfseries 01}
  (2019) 120} [\href{https://arxiv.org/abs/1811.07879}{{\ttfamily
  1811.07879}}].

\bibitem{Adler:1976zt}
S.L.~Adler, J.C.~Collins and A.~Duncan, \emph{{Energy-Momentum-Tensor Trace
  Anomaly in Spin 1/2 Quantum Electrodynamics}},
  \href{https://doi.org/10.1103/PhysRevD.15.1712}{\emph{Phys. Rev. D}
  {\bfseries 15} (1977) 1712}.

\bibitem{Adler:2004qt}
S.L.~Adler, \emph{{Anomalies to all orders}},  in \emph{{50 years of Yang-Mills
  theory}}, G.~'t~Hooft, ed., pp.~187--228 (2005),
  \href{https://doi.org/10.1142/9789812567147\_0009}{DOI}
  [\href{https://arxiv.org/abs/hep-th/0405040}{{\ttfamily hep-th/0405040}}].

\bibitem{Nielsen:1977sy}
N.K.~Nielsen, \emph{{The Energy Momentum Tensor in a Non-abelian Quark Gluon
  Theory}}, \href{https://doi.org/10.1016/0550-3213(77)90040-2}{\emph{Nucl.
  Phys.} {\bfseries B120} (1977) 212}.

\bibitem{Tarrach:1981bi}
R.~Tarrach, \emph{{The renormalization of FF}},
  \href{https://doi.org/10.1016/0550-3213(82)90301-7}{\emph{Nucl. Phys.}
  {\bfseries B196} (1982) 45}.

\bibitem{Collins:1984xc}
J.C.~Collins, \emph{{Renormalization}: {An Introduction to Renormalization, The
  Renormalization Group, and the Operator Product Expansion}}, vol.~26 of
  \emph{Cambridge Monographs on Mathematical Physics}, Cambridge University
  Press, Cambridge (1986),
  \href{https://doi.org/10.1017/CBO9780511622656}{10.1017/CBO9780511622656}.

\bibitem{Metz:2020vxd}
A.~Metz, B.~Pasquini and S.~Rodini, \emph{{Revisiting the proton mass
  decomposition}},  \href{https://arxiv.org/abs/2006.11171}{{\ttfamily
  2006.11171}}.

\end{thebibliography}

\providecommand{\href}[2]{#2}\begingroup\raggedright\endgroup

\end{document}